\documentclass[%
 reprint,
superscriptaddress,
 amsmath,amssymb,
 aps,
 showkeys,
]{revtex4-2}

\usepackage{graphicx}
\usepackage{dcolumn}
\usepackage{bm}
\usepackage{amsmath}
\usepackage[hidelinks]{hyperref}
\usepackage[all]{hypcap}

\DeclareMathOperator{\sech}{sech}

\begin{document}

\preprint{APS/123-QED}

\title{A sweeping twist defect as a topological flagellum \\ that drives colloid motion}

\author{Qi Xing Zhang}
\thanks{These authors contributed equally to this work.}
\affiliation{
Department of Chemical and Biomolecular Engineering,
University of Pennsylvania, Philadelphia, PA 19104, USA
}

\author{Claire Dor{\'e}}
\thanks{These authors contributed equally to this work.}
\affiliation{
Department of Chemical and Biomolecular Engineering,
University of Pennsylvania, Philadelphia, PA 19104, USA
}

\author{Mojtaba Rajabi}
\affiliation{
Department of Chemical and Biomolecular Engineering,
University of Pennsylvania, Philadelphia, PA 19104, USA
}
\affiliation{
Advanced Materials and Liquid Crystal Institute,
Kent State University, Kent, OH 44242, USA
}

\author{Edward B. Steager}

\affiliation{
Department of Mechanical Engineering and Applied Mechanics,
University of Pennsylvania, Philadelphia, PA 19104, USA
}

\author{Kathleen J. Stebe}
\email{kstebe@seas.upenn.edu}
\affiliation{
Department of Chemical and Biomolecular Engineering,
University of Pennsylvania, Philadelphia, PA 19104, USA
}

\date{\today}

\begin{abstract}
Nematic liquid crystals can dramatically reconfigure under dynamic forcing, providing exciting opportunities in active matter. Here, we study a hybrid disk colloid rotated by an external field which generates a dynamic companion topological defect. The disk moves faster when the defect sweeps across the disk’s face. We identify the defect as a non-singular twist wall, characterize the twist energy landscape, and identify the sweeping motion as a topological instability. As the defect sweeps, it reverses the handedness of twist and lowers the free energy in the fluid in the gap above the disk. Landau--de Gennes modeling shows that the sweeping wall behaves as a propagating director texture: the director field is nearly stationary in the wall frame, while nematogens rotate locally as the wall passes. The nematogens' rotation generates a viscous stress on the surface of the disk that hastens its propulsion. Thus, the defect acts as a  flagellum that powers colloid swimming, providing an example of a dissipative topological structure whose dynamics can be harnessed to perform useful work.

\end{abstract}

\keywords{Driven nematic colloids $|$ Topological solitons $|$ Backflow $|$ Instability $|$ Microswimmers}

\maketitle

\vspace{-1em}

\section{\label{sec:level1}Introduction}

Nematic liquid crystals (NLCs) provide an accessible, controllable  platform for studying topological defect dynamics \cite{ChuangCITL}. A colloid embedded in an NLC carries a companion topological defect whose structure is constrained by topology and selected by elasticity, anchoring, and particle geometry \cite{StarkPOCD, MusevicLCC, SmalyukhLCC}. When such a colloid is driven externally — by rotating spheres or disks \cite{RajabiSTFS, YaoTDPS}, or reciprocally forcing spheres or pulsating bubbles \cite{KimSPBS} — or internally, as in active droplets \cite{LavrentovichACIL, RajabiDSLO}, it translates under forcing that would not generate displacement in an isotropic fluid in the absence of inertia. Such propulsion relies on broken symmetries in the nematic orientational order, described by the director field \(\mathbf{n}\), a unit vector representing the local average molecular alignment, even when the forcing itself is symmetric or reciprocal \cite{RajabiSTFS}. In these scenarios, the companion defect plays a passive role: it lives near the particle and sets the symmetry of the surrounding director, but its own dynamics does not power propulsion. 
The question of whether a transient or dynamic defect can act as a propulsion engine in its own right — a topological flagellum — has remained largely unexplored \cite{AyaKOMS, SohnDOTS,DasPSIL,TangGAPO,LiEDTD, AghaCADN, BahrFIDL, CoparMCOT, GiomiCTBT}. 

In prior work, our group studied nickel-treated colloidal disks immersed in an NLC and rotated by an external magnetic field \cite{YaoTDPS}. One disk imposed uniform homeotropic anchoring, whereas the other had hybrid anchoring, with a planar top face and homeotropic sides and base. Both disks translated as they rotated, but only the hybrid disk exhibited a striking reconfiguration of its companion defect: the defect swept across the planar face in a swimming stroke, during which the disk moved faster. The translation of rotating disks can be explained by mechanisms recently introduced for a continuously rotated spherical colloid. Although lubrication effects and reduced effective viscosity in regions of reduced order were proposed as possible mechanisms, the nature of the defect and its role in propulsion remained unknown.

Here, we study the rotating hybrid disk to determine the topology of its companion defect, the instability that drives its swimming stroke, and its role in propulsion. We show that the defect is a twist wall that becomes unstable beyond a critical disk angle and then propagates across the disk as a soliton-like wave. Its motion is accompanied by molecular reorientation that generates material flow and anisotropic viscous stresses, thereby enhancing translation. After sweeping, the defect contracts along the disk edge and releases stored twist elastic energy, driving further disk rotation even after the external field is removed. Through this cycle of sweeping and contraction, the defect acts as a topological flagellum whose nonreciprocal motion propels the colloid, suggesting strategies for designing soft machines based on driven topological defects.

\section{\label{sec:level2}Results and discussion}

Our disk colloids are short right circular cylinders with height  $H$=30 µm and diameter  $2R$=75 µm; the finite height of the vertical sides plays a significant role in the defect structure and its dynamics. Hybrid disk colloids are fabricated following standard photolithographic methods using SU-8 epoxy resin on a silicon wafer; the disks are then sputtered with a 15 nm layer of nickel to make them ferromagnetic. The exposed surfaces of the disks are subsequently treated with the surfactant N-dimethyl-n-octadecyl-3-aminopropyl-trimethoxy silyl chloride (DMOAP) to impose homeotropic anchoring. During this process, the disks' bases are attached to the substrate. Thus, when the disks are removed from the substrate, the bases are uncoated and untreated. As a result, these disks have hybrid anchoring, with homeotropic anchoring on the tops and sides, but degenerate planar anchoring on the SU-8 base. All-homeotropic disk colloids are fabricated via a different protocol; essentially nickel-coated disks are first released from the substrate before receiving treatment to impose homeotropic anchoring over their entire surfaces (see \hyperref[sec:methods]{Materials and Methods} and SI Fig. \ref{SI:schematic} for details). 

Disks are dispersed in 4-cyano-4$'$-pentylbiphenyl (5CB) and confined between two glass slides imposing uniform planar anchoring, with a gap thickness $H_\text{gap}=50$–$60$ µm. After loading, the system is quenched into the nematic phase. The cell is placed on an inverted microscope and surrounded by four electromagnetic solenoids that generate an in-plane rotating magnetic field $\mathbf{B}$ of period $T$, which drives disk rotation. Throughout the paper, we adopt a coordinate convention in which the SU-8 face of hybrid disks is drawn on top; accordingly, ``above the disk'' denotes the gap adjacent to the SU-8 face. Counterclockwise (CCW) rotations are defined as positive and clockwise (CW) rotations as negative. We ensured that the rotating field does not reorient the nematic (no detectable Fredericks response at $|\mathbf{B}|\approx 10$~mT \cite{KedzierskiOMFD,EttingerMFDD}) and that residual magnetic-field gradients are too small to account for the observed translations. Experimental details are provided in \hyperref[sec:methods]{Materials and Methods}.

\subsection{\label{sec:sublevel1}Defect propelled swimming in NLC}

\begin{figure*}
\includegraphics{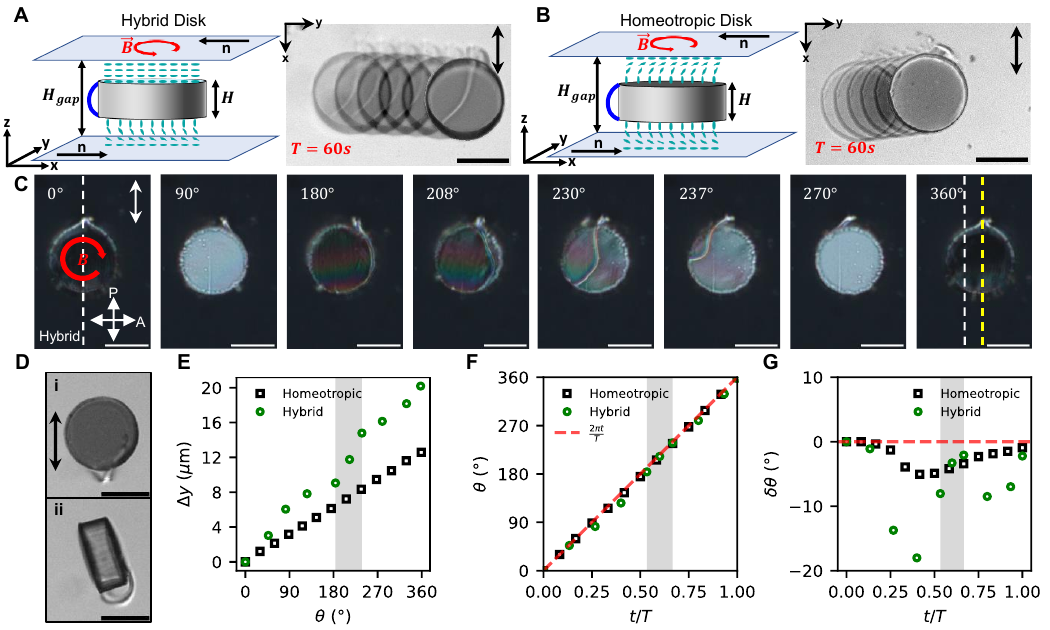}
\caption{Swim-stroke propulsion of nematic colloids. (A, B) For the hybrid disk (A) and the homeotropic disk (B): schematic of the director and defect structure (left) and bright-field overlays taken at equal time intervals ($\Delta t = T = 60$ s) over multiple periods (right). (C) Cross-polarized images of the hybrid disk at selected rotation angles $\theta$ over one full revolution ($T = 45$ s). White and yellow dashed lines mark the initial and final center-of-mass (COM) positions, respectively. (D) Bright-field images of a hybrid disk: i) at rest and ii) tilted with a hand-held magnet to reveal the companion defect. (E-G) For the hybrid disk in (C), the gray region indicates the swimming-stroke time interval; (E) disk displacement versus rotation angle $\theta$; (F) $\theta$ versus normalized time $t/T$ over one period; (G) $\delta\theta$, the phase lag between $\theta$ and $2\pi t/T$. Scale bars, 50 $\mu\mathrm{m}$.}
\label{fig:fig1}
\end{figure*}

To form the structures of interest, we rotate the ferromagnetic disks in an external magnetic field for a few full rotations and then remove the field. After this initial transient, similar dipolar defects form on both hybrid and homeotropic disks. The defect is located at one of the poles of the disk aligned with the far-field director, as shown schematically in blue in Figs.~\ref{fig:fig1}A and B. The equilibrium dipolar configuration of the hybrid disk is shown in Fig.~\ref{fig:fig1}D. When the microscope is focused on the SU-8 face (Fig.~\ref{fig:fig1}D i), the defect appears loop-like and attached to the edge of the disk adjacent to that face. To reveal its out-of-plane structure, we tilt the disk using a permanent magnet; the defect is then seen to attach to the opposite edge as well, forming a structure akin to a ``coffee-cup handle'' (Fig.~\ref{fig:fig1}D ii). The disks' hybrid anchoring, together with its sharp edges, departs from the smooth homeotropic case in which standard dipolar or Saturn-ring configurations are observed \cite{ConradiJNC, SenyukEPAT}.
Under a rotating field, however, the defects on hybrid and homeotropic disks behave very differently, as illustrated for CW disk rotation. On hybrid disks, the defect undergoes a complex reconfiguration: one end of the dipole elongates along the east edge of the disk while the other remains at the pole. Beyond a critical rotation angle, the elongated defect peels away from the edge and sweeps across the SU-8 face, as shown in the bright-field overlaid image in Fig.~\ref{fig:fig1}A and Movie S1. After this sweeping event, the defect contracts along the west edge and recovers its initial configuration at the pole. By contrast, defects on rotating homeotropic disks remain near the pole and are only weakly displaced by the viscous torque exerted by the rotational flow on the director, as shown in Fig.~\ref{fig:fig1}B and Movie S2. No defect elongation or sweeping is observed. The translational motion also differs strongly: both disks move periodically, but hybrid disks travel farther than homeotropic disks over each cycle. This suggests that the defect sweeping contributes to translation; for this reason, we refer to this reconfiguration as a ``swimming stroke''.

Insight into the origin of this swimming stroke comes from imaging the rotating disks between crossed polarizers. For the hybrid disk, the region above the SU-8 face shows a strong periodic modulation with the disk rotation angle $\theta$: it is nearly dark at $0^\circ$, becomes bright at $90^\circ$, returns near extinction at $180^\circ$ but now with faint interference colors, and brightens again at $270^\circ$ (Fig.~\ref{fig:fig1}C). 
This optical response is consistent with a twist developing in the thin gap between the SU-8 face and the lid, which rotates the polarization of transmitted light. At $0^\circ$ the director is uniform and the polarization, set parallel to it, is extinguished by the crossed analyzer. At $180^\circ$, $\pi$-twist rotates the polarization imperfectly, producing interference colors.

 The presence of twist above the SU-8 face is initially surprising because untreated SU-8 provides degenerate planar anchoring; we find that quenching from the isotropic to the nematic phase imprints an oriented planar easy axis on the SU-8 face aligned with the lid, and this oriented anchoring persists throughout the nematic phase (see SI Sec.~\ref{sec:SI_anchoring}, SI Figs. \ref{SI:rotate90andback}--\ref{SI:rotate90andquench}). By contrast, the homeotropic disk remains dark upon rotation (SI Fig. \ref{SI:homeotropicXPOL}), as expected for a splay--bend configuration in the \(xz\) plane (SI Fig. \ref{SI:rotatewithhomeotropic}).
 We hypothesize that the swimming stroke arises to relax the twist stored in the thin gap above the disk. Consistent with this idea is the absence of the swimming stroke for the homeotropic disk, which cannot generate twist in the gap.

We examine the disk dynamics in greater detail over one period of rotation in Figs.~\ref{fig:fig1}E--G. For the homeotropic disk, the displacement $\Delta y$ varies linearly with $\theta$, whereas for the hybrid disk the slope steepens abruptly during defect sweeping, which occurs over rotational angles $183^\circ$ to $238^\circ$ (shaded in gray). 
The translation of rotating homeotropic disks can be understood from mechanisms recently introduced for continuously rotated spherical colloids. In a dipolar configuration, the companion defect defines a polar axis around the colloid. Axisymmetric rotation couples through Leslie–Ericksen viscous stresses to this polar director field, producing a net viscous force and a displacement proportional to the rotation angle.
The hybrid disk's swimming stroke changes this scenario: elastic energy released during the sweeping event is associated with greater displacement. The effect of the defect is also evident in the evolution of the disk rotation angle, shown in Fig.~\ref{fig:fig1}F, where the diagonal red dotted line indicates the angular position of an object rotating uniformly at rate $2\pi/T$. The distinct responses are summarized in Fig.~\ref{fig:fig1}G, where the phase lag $\delta\theta$ between the disk rotation angle $\theta$ and $2\pi t/T$ is plotted. The lag remains small for the homeotropic disk, less than $5^\circ$, but reaches up to $20^\circ$ for the hybrid disk, indicating more complex elastic and viscous torques. The aim of this paper is to elucidate the nature of this sweeping defect and its role in enhancing disk translation.

\subsection{\label{sec:sublevel2}Simulation of the equilibrium defect structure}
The director structure is modeled within Landau--de Gennes theory, which accounts for topological defects; full equations, parameter values, and numerical details are provided in \hyperref[sec:methods]{Materials and Methods} and SI text Sec. \ref{sec:Landau_de_Gennes}. Nematic order is described by the \(\mathbf{Q}\)-tensor, a symmetric traceless \(3\times3\) matrix whose leading eigenvalue and eigenvector define the scalar order parameter $S$ and director $\mathbf{n}$. The total free energy 
\begin{equation}
\begin{aligned}
F_\text{tot}[\mathbf{Q}]
&=
\int_{\Omega}
\left[
f_{\text{bulk}}(\mathbf{Q})
+
f_{\text{elastic}}(\mathbf{Q}, \nabla \mathbf{Q})
\right]
\, \mathrm{d}V
\\
&\quad
+
\int_{\partial \Omega}
f_{\text{surf}}(\mathbf{Q})
\, \mathrm{d}S,
\end{aligned}
\end{equation}
comprises a bulk phase term favoring nematic order, an elastic term penalizing distortions, and a surface term imposing anchoring. To focus on the director field topology, flow is neglected and \(\mathbf{Q}\) evolves by relaxational dynamics. We use a two-constant approximation, $L_1=6$pN and $L_2=12$pN (twist/splay ratio $\frac{K_{22}}{K_{11}}=0.5$), to capture the lower twist cost of 5CB while avoiding nonlinear elastic terms. The simulation reproduces the experimental hybrid disk geometry, with planar anchoring along \(\mathbf{e}_x\) on the confining plates, planar anchoring along \(\mathbf{e}_{\theta}=(\cos\theta,\sin\theta,0)\) on the disk top face, and homeotropic anchoring elsewhere on the disk.

Before simulating the quasistatic rotation of the hybrid disk, we seek to recover an equilibrium configuration similar to the experimental dipole state (Fig. \ref{fig:fig1}E) and set $\theta=0$.
For reasons that will become clear later, we choose a nontrivial initial director field in which the mismatch between the radial anchoring at the disk’s cylindrical wall and the uniform planar anchoring at the substrates is accommodated by a twist deformation. This twist grows from $0^\circ$ at the pole opposite the dipolar defect to $\pm180^\circ$ at the defect location (see SI Sec. \ref{sec:SI_companion} for details). After relaxation, the director reaches an equilibrium state where the dipole defect is a non-singular handle-shaped region connecting to the top and bottom edges of the disk. Within this structure, the director escapes vertically, with a smooth transition between the twist domains with opposite handedness on either side of the defect (see Fig. \ref{fig:fig2}A). The region where the vertical tilt of the director is $|n_z|>0.9$ is colored in orange. Because the director escapes upward when we look at the handle from the front (Fig. \ref{fig:fig2}A ii), we refer to this specific dipole-state as the ``coffee-cup handle escaping upward'' (CCHU);  we show in SI Fig. \ref{SI:companion_defects} D, a similar dipole state in which the director escapes downward in the handle (CCHD). We stress that no singularity is present in the bulk for CCHU, making our companion defect entirely non-singular, unlike the configurations discussed in the literature for hybrid and faceted structures \cite{ConradiJNC}. This equilibrium non-singular companion defect is made accessible by using a topologically non-trivial initial configuration. Spatial variations of the optical indices in the handle would act as a microscopic lens, rendering the structure visible under bright field microscopy, consistent with experiment.

\subsection{\label{sec:sublevel3}Simulation of a rotated disk that eliminates twist via sweeping of a singular twist defect}

\begin{figure*}
\includegraphics{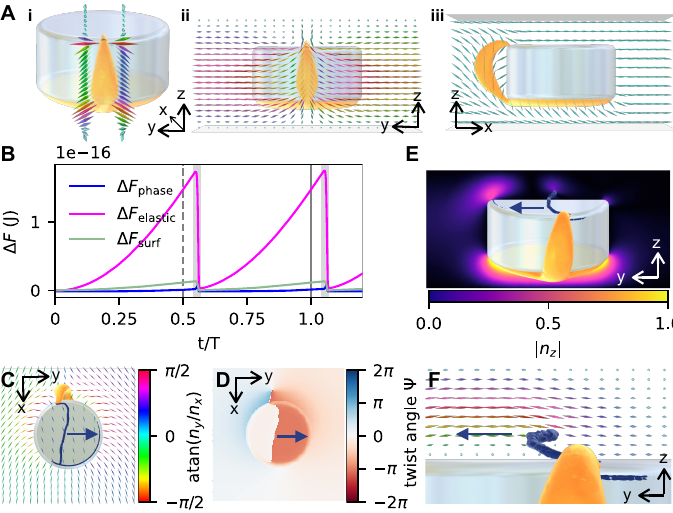}
\caption{Elimination of twist by sweeping singular disclination lines twice per rotation. (A) CCHU state, similar to the experimental dipole. Regions with strong vertical director tilt ($|n_z|>0.9$) are shown in orange, and singular cores ($S<0.7$) in dark blue. (B) Bulk phase, elastic, and surface free energies during quasistatic rotation. Gray intervals denote the sweeping of a singular defect across the planar disk face; dashed and solid vertical lines mark half and full rotations, respectively. (C) Director field around the disk in the midplane $z=0$ and on the planar disk face at $\theta=200^\circ$. (D) Corresponding map of the total twist from $z=0$ to the top lid at $z=H_{\mathrm{gap}}/2$. (E) Vertical director tilt in the $yz$ midplane during sweeping at $\theta=200^\circ$, showing that the director remains planar in the gap. (F) Director structure around the singular twist defect. Simulations use $L_1=6$ pN and $L_2=12$ pN; panels (B-F) show the first rotation. In (A), (C) and (F), the color scale indicates the polar angle of the director projected onto the $xy$ plane. In (C-F), the dark blue arrow represents the sweeping direction of the defect.}
\label{fig:fig2}
\end{figure*}

To simulate the quasistatic rotation of the disk absent material flow, we subject the CCHU state to a time-dependent CCW rotation of the anchoring boundary condition on the top of the disk with $\theta(t)=+2\pi t /T$, with $T=100$ ms. The rotation period $T$ is selected to be much greater than the elastic relaxation time of the simulation $\tau_\text{elastic}=\gamma R^2/K \approx 2$ ms so that the NLC remains in equilibrium until it reaches an unstable point. During the rotation, we track the variation of the phase, elastic and surface free energy of the NLC (Fig. \ref{fig:fig2}B) as well as the net twist angle accumulated across the upper half-cell $\Psi(x,y)=\int_{0}^{L_z/2} \mathbb{I}_{\Omega}(\mathbf{r})\partial_z\phi(\mathbf{r}) \text{d}z$ where $\phi$ is the azimuthal angle of the director, and $\mathbb{I}_{\Omega}$ is the indicator function of the NLC domain (Fig. \ref{fig:fig2}D). In the first half of the rotation, the elastic energy increases with time as a CW twist domain grows in the gap above the disk. The system becomes unstable at $\theta \approx 199^\circ$, when a singular disclination line unpins from the top edge and sweeps over the disk's top surface (Movie S3). This first sweeping event is shown in Figs. \ref{fig:fig2} C-F; the core of the disclination ($S<0.7$) is colored in dark blue. The sweeping event reduces the total twist by $-\pi$. Thus, when the sweeping is complete, the gap above the disk has recovered a nearly untwisted structure (Figs. \ref{fig:fig2} D and F), with an abrupt drop in elastic free energy. The director winds by $|\pi|$ about the vertical axis around the sweeping defect line (Fig. \ref{fig:fig2}F), further confirming that this sweeping defect is a twist disclination line \cite{NehringCOTS,WangAWOD}. The director remains planar in the gap while the defect sweeps (Fig. \ref{fig:fig2}E). Since twist is erased by the sweeping disclination line after half a rotation of the disk, the dynamical period of the director is half of the period of rotation, and we count two sweeping events per full rotation of the disk (Fig. \ref{fig:fig2}B). The imposition of twist by rotation of a boundary and its erasure mediated by a singular twist disclination line is related to the work of Long et al. \cite{LongFRMI}, however, in their work the twist disclination line is already present in the bulk, while in our case it nucleates along the disk's edge as the colloid rotates.

Consistent with our hypothesis, this simulation predicts that defect sweeping relaxes the twist stored in the gap. However, two experimental features rule out this singular-disclination scenario: (i) only one sweep is observed per rotation, rather than two, and (ii) the interference colors on either side of the sweeping defect (Fig. \ref{fig:fig1}C) reveal that twist persists on both sides of the defect, rather than being eliminated. The singular-disclination mechanism therefore does not describe the swimming stroke, but it captures transient dynamics relevant to dipole-defect formation, which we examine next.

\subsection{\label{sec:sublevel4}Birth of dipole: transition from quadrupolar to dipolar defect}

\begin{figure*}
\includegraphics{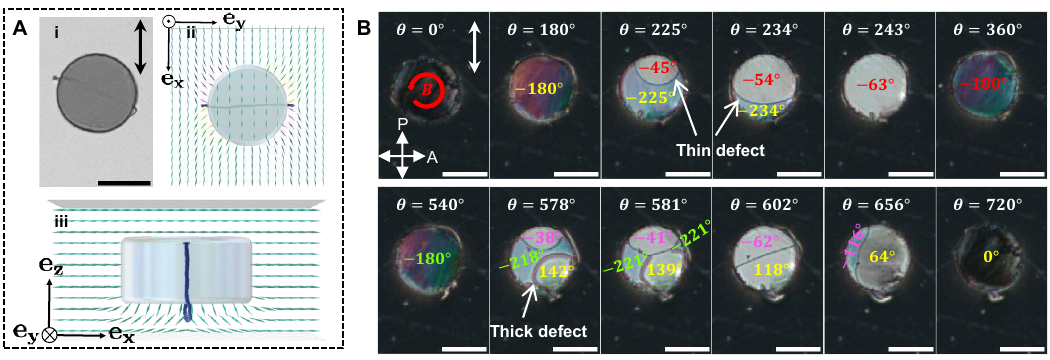}
\caption{Birth of the dipole for the hybrid disk. (A) Initial quadrupolar state obtained after quenching to the nematic phase: i) bright field microscopy image in which singular disclination lines are visible on the sides and across the homeotropic face of the disk, ii) top view and iii) side view of the quadrupolar state in simulations in which the core of the singular defect line ($S=0.7$) is colored in dark blue. (B) Time sequence of cross-polarized images showing typical steps of a transition from quadrupole to dipole. The transition from $\theta = 360^\circ$ and $\theta=540^\circ$ is not shown. Twist domains are identified and labeled with a color-code. Scale bars, 50 $\mu\mathrm{m}$.}
\label{fig:fig3}
\end{figure*}

To clarify the discrepancy between the simulation and the experimental behavior, we examine the transient dynamics leading to the dipolar state. The hybrid disk and the NLC are loaded into the cell in the isotropic state. Immediately after quenching, the NLC around the disk assumes a quadrupolar arrangement, with two fragments of defect evident in a top view on either side of the disk, aligned on the axis perpendicular to the far field director. In simulations, the quadrupolar state is obtained after relaxation from a uniform director parallel to the $x$-axis (Figs. \ref{fig:fig3}A ii and iii). A half-Saturn ring configuration is revealed in which a single disclination line connects the end points on either side of the disk. The core of the singular defect, shown in dark blue  in the simulation, carries a $-1/2$ winding number throughout the disclination. 

Rotation of the disk under the rotating magnetic field destabilizes the initial quadrupolar defect configuration and drives a topological transition to the dipolar state occurring over $k$ half-rotations via intermediate transitions. An example is provided in Fig.~\ref{fig:fig3} and Movie S4. The transition proceeds via the sweeping of two distinct types of twist defects as identified by Nehring \cite{NehringCOTS}: (i) singular twist disclination lines, which have a singular core where the director is undefined and separate domains whose twist differs by $\pi$; these appear as thin dark lines in optical microscopy; and (ii) non-singular twist walls, in which the director remains well defined and the twist changes by $2\pi$ across the wall. Twist walls relax twist by allowing out-of-plane splay and bend, and appear as thick bands due to the smooth vertical director tilt, which acts as a weak diverging lens \cite{NehringCOTS}.

A typical quadrupolar to dipolar transition imaged between crossed-polarizers is shown in Fig. \ref{fig:fig3}B versus the disk's rotation angle $\theta$. The director progressively twists until $\theta \approx 225^\circ$, where a thin disclination line sweeps from top to bottom, reducing twist by $-\pi$ as attested by the lack of interference colors in the newly formed domain. Singular sweeps of this kind recur every half rotation until $\theta \approx 540^\circ$ (not shown). Then, a thick twist wall and a singular disclination enter simultaneously from opposite poles; they recombine, and one further singular sweep locks the system into the dipolar state, with the coffee-cup handle anchored at the south pole. Thereafter, a thick defect sweeps across the disk once per rotation, as in the experimental swimming stroke dynamics reported in Fig.~\ref{fig:fig1}. We identify this thick sweeping defect as a twist wall across which the director twist changes by $2\pi$. Because it sweeps the disk near $\theta \approx \pi$, it effectively reverses the twist handedness in the gap above the disk. 

The observation that the dipole defect nucleates as a twist wall and remains continuously connected to the sweeping twist wall sheds light on the coffee-cup-handle structure seen in experiments (Fig.~\ref{fig:fig1}D) and guided the ansatz used in the simulation (Fig.~\ref{fig:fig2}A). The homeotropic vertical side of the disk imposes a radial director orientation in the cell mid-plane, requiring twist to match the oriented planar anchoring on the lid and bottom plate. The twist in the lower half of the cell mirrors that in the upper half, yielding zero net twist overall. In the dipolar state, the twist deformation grows from \(0^\circ\) at the pole opposite the defect to \(\pm 180^\circ\) at the defect location, with opposite handedness on the west and east sides. The handle defect begins and ends on twist walls emerging from the disk's top and bottom edges, providing a smooth transition between \(\pi\) CW and \(-\pi\) CCW twist domains at the pole. The rest of the ``coffee-cup handle'' is a vertically escaped region connecting these two twist walls. In simulations, this connection forms when the director escapes in the same direction in both twist walls (SI Sec.~\ref{sec:SI_twist_walls} and SI Figs.~\ref{SI:twist_walls_connection}--\ref{SI:twist_walls_no_connection}).

We highlight that the squat cylindrical structure of the hybrid disk, together with its planar surface, encourage the director to be planar everywhere except  below the homeotropic face. Therefore, we expect twist to be the key distortion imposed by the hybrid disk, unlike a homeotropic sphere around which the director is more three dimensional \cite{ConradiJNC}. The distribution of twist around the static disk for the quadrupolar and dipolar states is discussed further in SI Sec.~\ref{sec:SI_companion}. Homeotropic disks also form a quadrupolar state immediately after quenching, and transition to a dipolar structure upon rotation via defect rearrangement on the vertical sides of the disk (SI Fig. \ref{SI:homeotropictransition}); however, no defect sweeping occurs as there is no twist above the disk. We show numerically that homeotropic disks can support a similar dipole defect (SI Fig. \ref{SI:companion_defects} E).    

\subsection{\label{sec:sublevel5}Simulation of rotated disk with a sweeping non-singular twist wall}

\begin{figure*}
\includegraphics{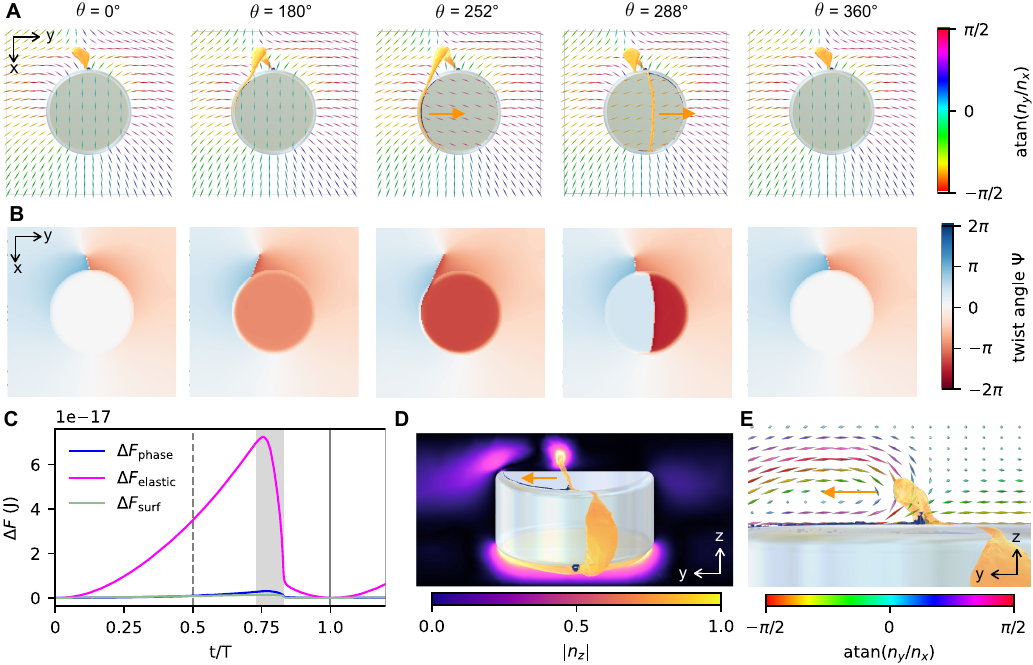}
\caption{Elimination of twist by sweeping a non-singular twist wall once per rotation. (A)  Director field around the disk at $z=0$ and on the planar disk face over one full rotation. (B) Corresponding map of the total twist from $z=0$ to the top lid at $z=H_{\mathrm{gap}}/2$. (C) Bulk phase, elastic, and surface free energies during quasistatic rotation. The gray shaded region denotes sweeping of the non-singular twist wall across the planar disk face; dashed and solid vertical lines mark half and full rotations, respectively. (D) Vertical director tilt in the $yz$ midplane during sweeping at $\theta=288^\circ$, showing a vertical director core within the twist wall. (E) Director structure around the non-singular twist wall. Simulations use $L_1=1$ pN and $L_2=12$ pN. In A, D, and E, the orange arrow represents the sweeping direction of the defect.}
\label{fig:fig4}
\end{figure*}

We revisit quasistatic simulations of disk rotation to recover the sweeping twist wall seen in experiments.
Three-dimensional Landau--de Gennes simulations are necessarily performed at $\sim 1$ µm scale, while experiments take place at $10-100$ µm scale. This difference in scale changes the relative cost of non-singular walls and singular defect cores. In particular, at these smaller scales, the cost of elastic distortions around singular defect lines is energetically inexpensive. Consistent with this, the simulation using standard 5CB elastic constants relaxes the twist in the gap by the sweeping of a singular twist disclination line. 
To access the non-singular twist wall topology observed in the experiment, we allow the system to twist further before the onset of the instability. This is achieved by lowering the cost of twist relative to splay and bend by setting $L_1=1$ pN and $L_2=12$ pN (yielding $K_{22}/K_{11}=0.14$), keeping all other parameters unchanged. Starting from the CCHU state, the director twists quasistatically in the gap until a critical angle $\theta_c\approx 270^\circ$, where the edge defect on the west rim becomes unstable, detaches, and sweeps across the disk face from west to east (Movie S5; Figs.~\ref{fig:fig4}A and B). This event releases elastic free energy (Fig. \ref{fig:fig4}C) as it relaxes the accumulated twist by $2\pi$, converting the overtwisted CW state into a weakly twisted CCW state; the remaining twist is then unwound as the disk completes the rotation (Fig. \ref{fig:fig4}B).

The simulation also clarifies the origin of ``pinning'' of the dipole defect. As the disk rotates, twist builds up in the thin gap above the disk with a handedness opposite to that of the twist surrounding the disk. The handle therefore elongates along the west rim, from the north to the south pole, to form the boundary between these two twist domains of conflicting handedness (Figs. \ref{fig:fig4}A and B). This twist partitioning mechanism explains the apparent pinning of the handle on the rim, without invoking surface roughness as a pinning site. When the edge defect becomes unstable, it detaches and sweeps across the disk face as a non-singular twist wall. The sweeping wall, defined as $\lvert n_z \rvert > 0.9$, is a tubular region in which the director escapes vertically in an otherwise planar cell (Figs.~\ref{fig:fig4}D,E). It separates two $\sim\pi$-twist domains of opposite handedness and accomplishes the reversal of handedness by rotating the director about an axis parallel to the wall, analogous to a N\'eel wall in a 2D ferromagnet \cite{MatzenDADW} (see SI Fig.~\ref{SI:twist_wall}). Once detached from the rim, the wall propagates with an essentially invariant internal structure, i.e. as a soliton-like traveling defect. The vertical escape direction of the wall (upward or downward) defines distinct topologies; in our simulations the sweeping wall always escapes downward, irrespective of the escape direction in the handle, which leads to a point singularity in the connection between the handle and the wall (Fig. \ref{fig:fig4}A). When the CCHD state is used instead, handle defect and sweeping defect connect continuously without point singularity (SI Fig.~\ref{SI:CCHD_rotation}).

\subsection{\label{sec:sublevel6}Swimming stroke as a topological instability}

\begin{figure*}
\includegraphics[width=0.85\textwidth]{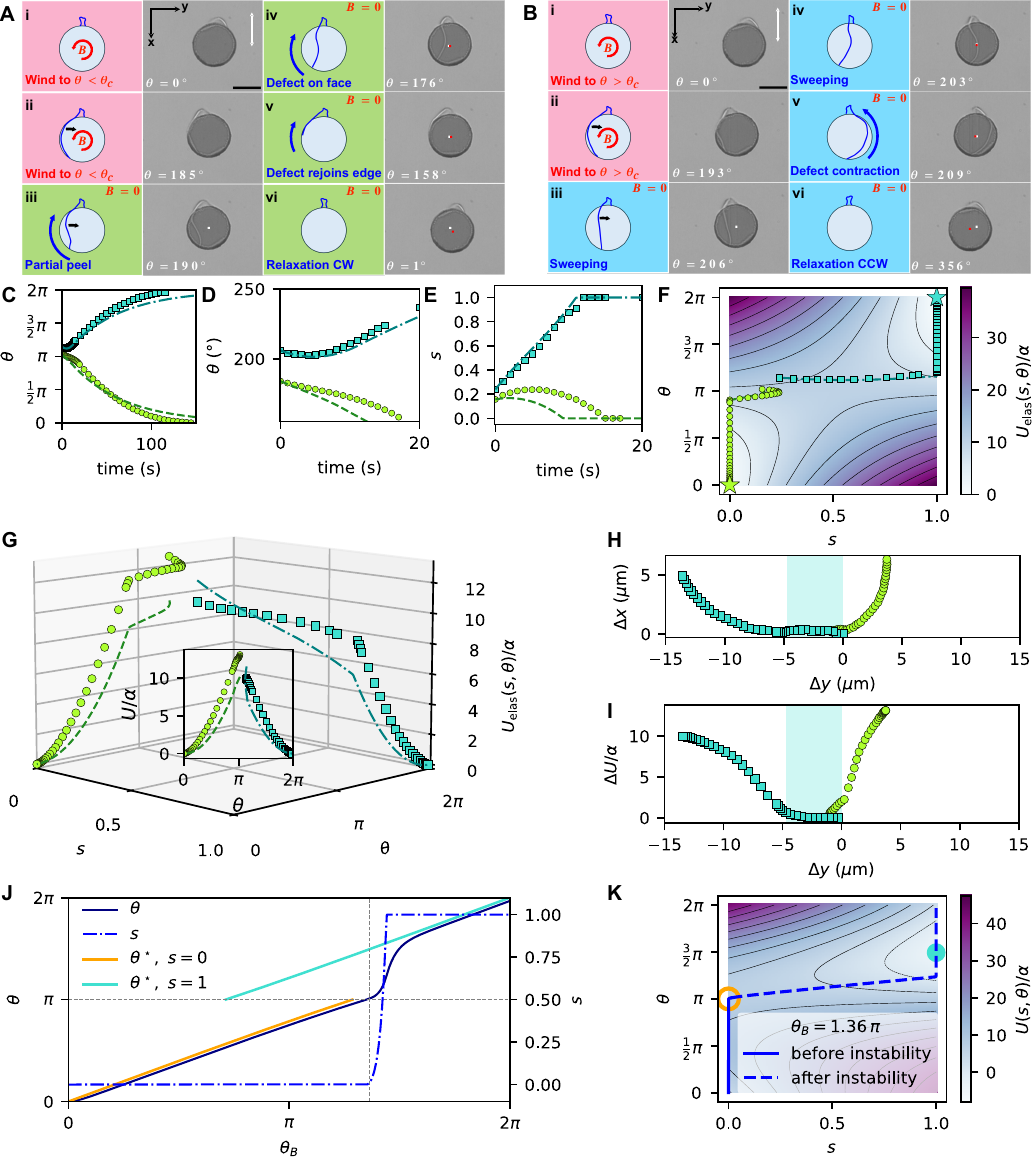}
\caption{Swim stroke as a topological instability. Schematics and corresponding experimental bright-field images of a hybrid disk rotated slowly to an angle (A) below and (B) above the critical angle (pink), then allowed to relax back to equilibrium after the external field is switched off (lime green and cyan blue, respectively). The disk COM is marked by a white dot at the instant the field is turned off and by red dots at later times. In the remaining panels, markers denote experimental data (green circles, subcritical; cyan squares, supercritical) and lines the model (green dashed, subcritical; teal dash-dotted, supercritical). (C) Disk rotation angle, $\theta$, during relaxation. (D,E) Coupled dynamics of $\theta$ and $s$ while the defect lies on the disk face. (F) Trajectories in the $(s,\theta)$ plane on the model elastic energy landscape, $U(s,\theta)$. (G) Inferred experimental elastic potential, $U_{\mathrm{exp}}(s,\theta)$, and model elastic energy as functions of $s$ and $\theta$; inset: energy projected onto the $\theta$ axis. (H,I) The turquoise shaded region indicates the defect-sweeping interval in the supercritical dynamics. (H) Horizontal and vertical displacements of the disk COM, $\Delta y$ and $\Delta x$. (I) Dissipated energy, $\Delta U/\alpha$, versus lateral displacement, $\Delta y$. (J,K) Simulated dynamics of the disk-defect system under a rotating magnetic field, with $a=10$, $b=40$, and $\mu=10$. (J) Evolution of $\theta$, $s$, and stable fixed points $\theta^\star$ as a function of magnetic-field orientation, $\theta_B$; the orange and turquoise branches correspond to stable fixed points on the boundaries $s=0$ and $s=1$, respectively. (K) Simulated trajectory in the $(s,\theta)$ plane. The colormap shows the total energy landscape at the onset of the instability; the stable fixed point on the $s=1$ boundary is marked by a turquoise disk, and the saddle point on the $s=0$ boundary by an orange circle. Scale bars in A and B, 50 $\mu\mathrm{m}$.}
\label{fig:fig5}
\end{figure*}

Very thin ferromagnetic disks that are rotated in NLCs have topological instabilities that feature dislocation loop shedding and rapid rearrangement of the disk and defect \cite{RovnerLehenyEHTC}. However, in these systems, unlike our finite thickness hybrid disks, directed motion upon defect shedding  was not reported. To understand the stability of our defect and its role in generating translation, we wind the hybrid disk slowly CCW using the external magnetic field rotating with rotational period $T=80 \ \text{s}$, larger than the relaxation time of the NLC. We then switch off the field at some rotation angle of the disk $\theta_\text{0}$ near $\pi$ and observe the system's relaxation dynamics, tracking its orientation $\theta$. 

\subsubsection{\label{subsec1}Sweeping reverses the twist handedness}
The system relaxes along distinct trajectories depending on the release angle $\theta_\text{0}$, as shown in Fig. \ref{fig:fig5}. For $\theta_0 \leq 187^\circ$, we observe that the disk relaxes by rotating CW, that is it returns to its equilibrium position following the same path it took during the winding phase Movie S6. For this reason, we refer to this perturbation as the reversible, or sub-critical, regime. For $\theta_0 \geq 193^\circ$, defect sweeping occurs, and the disk relaxes by a CCW rotation, that is it continues its rotation in the same sense as the winding phase Movie S7. Hence, we identify this perturbation as the irreversible, or supercritical regime. Schematics and optical microscopy images of the experiment are shown in  Figs. \ref{fig:fig5}A for the sub-critical dynamics with $\theta_0 = 187^\circ$ and B for the supercritical dynamics with $\theta_0 = 206^\circ$. In the schematics, pink corresponds to the winding regime, and in the entire Fig. \ref{fig:fig5}, lime green refers to the sub-critical relaxation, and cyan to super-critical relaxation. We identify $\theta_\text{0}$ ranging from $\sim 188^\circ-205^\circ$ as a zone in which rapid changes challenge our ability to capture disk and defect dynamics. Additional relaxation dynamics for $\theta_0 = 79^\circ, 137^\circ, 248^\circ, 281^\circ$ are provided in SI Fig. \ref{SI:elastic_relaxation}. The opposite disk rotations observed in the subcritical and supercritical regimes validate that twist wall sweeping reverses the handedness of the twist stored in the gap. 

\subsubsection{\label{subsec2}Coupled dynamics during defect instability}
When $\theta$ approaches $190^\circ$, the elongated defect undergoes an instability: it detaches from the west edge and initiates sweeping across the disk face, as seen in Figs. \ref{fig:fig5}A iii) and B ii) and iii). In the subcritical dynamics, immediately upon release and while the disk relaxes by CW rotation, the defect partially sweeps (Fig. \ref{fig:fig5}A iii and iv) before receding and repinning along the west edge (Fig. \ref{fig:fig5}A v). In the supercritical  regime, upon release, the disk rotates slightly CW before setting up rotation in CCW direction for good, meanwhile the twist wall irreversibly sweeps across the face of the disk, and then pins along the east edge (Fig. \ref{fig:fig5}B iii-v).
To capture quantitatively this coupling, we measure the progress of a sweep with a scalar variable $s\in[0,1]$, defined as $s=A_\mathrm{swept}/A_\mathrm{disk}$, the fraction of the projected disk area already swept by the defect. Thus $s=0$ and $s=1$ denote the pre- and post-sweep states, respectively.
The coupled evolution of $\theta$ and $s$ during the sweeping event is reported separately in Figs. \ref{fig:fig5}D and E, respectively, and the trajectories in the ${s,\theta}$ plane are displayed in Fig. \ref{fig:fig5}F.

To explain the coupled dynamics of the disk and the defect, we propose a model energy $U_\mathrm{elas}(s,\theta)$ that only accounts for the twist free energy in the gap of thickness $d$ above the disk. With the assumption that the wall marks a jump of $2\pi$ in the twist,

\begin{equation}
    U_\mathrm{elas}(s,\theta) = \alpha((1-s)\theta^2+s(\theta-2\pi)^2),
\end{equation}

with $\alpha = \frac{\pi R^2 K_{22}}{2d}$ the scale of twist free energy. The model energy landscape, featured in the colormap in Fig. \ref{fig:fig5}F, has a saddle-point at $\{s =\frac{1}{2},\ \theta=\pi\}$, and a unique minimum at $\{\theta = 0\ \text{mod}\ 2\pi, s=0\ \text{mod}\ 1\}$ marked by the stars. The dynamics being overdamped, the governing equations of $\theta$ and $s$ are

\begin{align}
    C_D \dot{\theta} &= -\frac{\partial U_\mathrm{elas}}{\partial \theta}, \label{overdamped_theta} \\
    C_W \dot{s} &= - \frac{\partial U_\mathrm{elas}}{\partial s} \label{overdamped_s}
\end{align}

where $C_D$ and $C_W$ are effective drag coefficients for the rotation of the disk and the motion of the defect, respectively. The characteristic relaxation time of the disk, $\tau_\mathrm{elas}= \frac{C_D}{2\alpha}$, is inferred from the dynamics  of $\theta$ outside of the sweeping event (SI Sec.~\ref{sec:SI_toy_model} and Fig. \ref{SI:elastic_relaxation}). We estimate $\tau_\mathrm{elas} \approx 59\ \text{s}$, giving $C_D \approx 5 \times 10^{-14} \text{J} \cdot \text{s}$ for a gap above the disk of $d=10$ µm. The linear growth of $s$ during the supercritical dynamics provides an estimate of $C_W=3.6 \times 10^{-14} \text{J}\cdot \text{s}$. Using the rescaled time $\tilde{t}=t/\tau_\mathrm{elas}$, the dimensionless equations

\begin{align}
    \dot{\theta} &= 2 \pi s - \theta, \\
    \dot{s} &= \mu(\theta-\pi)
\end{align}

determine the trajectory of the system via a unique parameter, $\mu = \frac{2 \pi C_D}{C_W}$. Clamping boundary conditions are applied to the second equation to ensure that $s$ stays in the interval $[0,1]$. The second equation shows that that the speed of the wall is proportional to the over-twist in the initial domain $\theta -\pi$, and that a twist angle $\theta$ greater than $\pi$ triggers sweeping.

We simulate the dynamics of our model with the same initial conditions $s_0,\ \theta_0$ upon release as in those in the experiment for the sub- and supercritical trajectories, using $\tau_D = 59 \ \text{s}$ and $\mu =10$. Our model produces trajectories that qualitatively match the experiment (Figs. \ref{fig:fig5}C-F). We do not expect quantitative prediction as our simple free energy model takes into account the energy of the twist domains and ignores other distortions, such as the splay/bend distortion associated to the twist wall, which depends on the wall shape, and the distortions around the disk that are also growing as the disk is rotated out of equilibrium. Moreover, the effective drag coefficients, constant in our model, are in principle time-dependent in the experiment, reflecting the changing organization of the NLC. These caveats  notwithstanding, the model captures key features and allows us to identify the essential mechanisms at play. In our experiments, we release the system of the disk and the defect wall in the vicinity of the saddle-point. The trajectory followed by the system to relax towards equilibrium is not only defined by the elastic energy landscape, but also by the relative drag exerted on the disk and the defect wall, captured by $\mu$ (SI Sec.~\ref{sec:SI_toy_model} and Fig. \ref{SI:elastic_relaxation}).

Using our estimates for $C_D$ and $C_W$, it is possible to infer the elastic energy of the system $U_\text{exp}(s,\theta)$ if we assume that the dynamics of $\theta$ and $s$ are governed by Eqs. \ref{overdamped_theta} and \ref{overdamped_s}. 

\begin{equation}
    U_\text{exp} (s(t),\theta(t)) = - \int_{t_0}^{t} (C_D\dot{\theta}^2+C_W\dot{s}^2)dt + \text{const.}
\end{equation}

where the constant is adjusted so that the energy is zero in the ground state $\{\theta = 0\ \text{mod}\ 2\pi, s=0\ \text{mod}\ 1\}$. The inferred energy along the experimental trajectories is plotted against $s$ and $\theta$ in Fig. \ref{fig:fig5}G, together with the energy of the simulated trajectories. Outside of the sweeping event ($s=0$ or $s=1$), the energy is quadratic in $\theta$. 

\subsubsection{\label{subsec3}Displacement of the disk during the relaxation}
During the relaxation process, the disk not only rotates, it also translates. In the images in Figs. \ref{fig:fig5}A and B, the position of the disk's COM is shown as a white point at the moment that the magnetic field is removed and as a red point in subsequent frames.  Displacements in the lab frame $\Delta y$ and $\Delta x$ are reported in  Fig. \ref{fig:fig5}H. The vertical displacement along $x$ is similar for the subcritical and supercritical trajectories ($\Delta x \approx 5$ µm), while the horizontal displacements are not only in opposite direction, they also differ in magnitude: $\Delta y \approx -15$ µm for the supercritical relaxation and  $\Delta y \approx 5$ µm for the subcritical relaxation. In fact, a third of the horizontal displacement of the supercritical dynamics is acquired during the sweeping of the defect. Moreover, the dissipated energy $\Delta U(t) = \int_{t_0}^{t} (C_D\dot{\theta}^2+C_W\dot{s}^2)dt$ is not a linear function of the displacement $\Delta y$, as seen in Fig. \ref{fig:fig5}I, and during sweeping the dissipated energy per unit of displacement is low, suggesting that the sweeping defect is particularly efficient at converting elastic free energy into translation. Translation continues after repinning of the defect along the edge, as the disk rotates to equilibrium, however, the energy expenditure per unit of displacement is 5 times higher in this regime.

\subsubsection{\label{subsec4}Disk under forced rotation}
The relaxation experiments discussed above reveal the elastic part of the landscape governing the coupled variables $(s,\theta)$. We now extend this model to the driven case, in which the in-plane magnetic field rotates at constant rate, with orientation $\theta_B(t)=2\pi t/T$. The total energy, combining magnetic and elastic contributions, is
\begin{equation}
    U(s,\theta)=\alpha\Big[-\kappa\cos(\theta_B-\theta)+(1-s)\theta^2+s(\theta-2\pi)^2\Big],
\end{equation}
where $\kappa=\mu_B B/\alpha$ compares magnetic and elastic torques.

Using the rescaled time $\tilde t=t/T$, the overdamped dynamics become
\begin{align}
\dot{\theta} &= b\sin(2\pi \tilde t-\theta)+a(2\pi s-\theta),\\
\dot{s} &= \mu a (\theta-\pi),
\end{align}
with clamping imposed to keep $s\in[0,1]$. Here, $a=T/\tau_\mathrm{elas}$, and $b=T/\tau_\mathrm{mag}$, where $\tau_\mathrm{mag}=C_D/\mu_B B$ is the response time associated with the magnetic torque.

Figs. \ref{fig:fig5}J and K and Movie S8 show the simulated dynamics over one forcing cycle for $a=10$, $b=40$, and $\mu=10$. In the slow-driving limit, $T\gg \tau_\mathrm{elas},\tau_\mathrm{mag}$, the system evolves quasi-statically near stable fixed points of the instantaneous driven landscape. These stable equilibria lie on the boundaries $s=0$ and $s=1$, yielding two stable branches $\theta^\star(\theta_B)$ shown in orange and turquoise in Fig.~\ref{fig:fig5}J. A fuller bifurcation analysis of the driven landscape is provided in the SI Sec.~\ref{sec:SI_toy_model} and  Fig. \ref{SI:bifurcation_diagrams}. Fig. \ref{fig:fig5}J also shows the simulated evolution of $\theta$ and $s$ as functions of $\theta_B$. As the field rotates, the system tracks the $s=0$ branch until this boundary equilibrium loses stability in the $s$ direction at $\theta=\pi$. This triggers a rapid, non-equilibrium jump to the competing stable boundary branch $s=1$, which the system follows for the remainder of the cycle. Over a range of $\theta_B$, both boundary states coexist, reflecting the bistability of the unswept ($s=0$) and swept ($s=1$) configurations. Fig. \ref{fig:fig5}K offers a geometric view of the simulated trajectory in the $(s,\theta)$ plane, superimposed on the energy landscape at the onset of the instability. Sweeping is therefore an instability-driven switching event between the bistable states $s=0$ and $s=1$. In experiment, defect sweeping occurs at a critical angle slightly larger than $\pi$. We attribute this delay to kinetic barriers, which the model captures via a pinning term (SI Fig. \ref{SI:pinning_energy}). 

In summary, slow rotation stores twist energy in the gap above the disk; past a critical angle, a sweeping twist wall releases this energy in a topological instability that reverses the twist handedness. The disk translates substantially during the sweep itself, when its orientation barely changes, making this regime particularly effective at converting elastic energy into translation. How the wall generates the propulsive stresses responsible for this motion is the focus of the next section.

\subsection{\label{sec:sublevel7}Hydrodynamics of a translating twist wall}

\begin{figure}[b]
\includegraphics[width=0.45\textwidth]{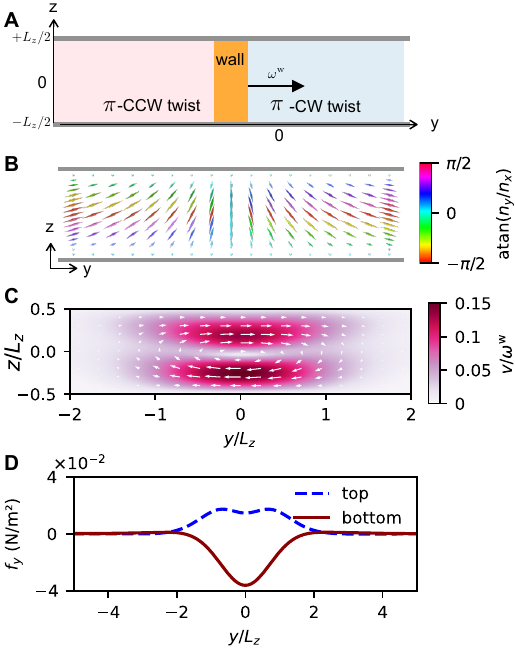}
\caption{Hydrodynamics of a translating twist wall (A) Schematic of a twist wall separating a $\pi$-CCW twist domain for $y<0$  from a $\pi$-CW twist domain for $y>0$. The wall translates with velocity $\omega^\mathrm{w}$ towards $+y$. (B) Director structure of a twist wall with upward escape. (C) Velocity field generated by wall translation. (D) Profile of the surface force density on the top and bottom substrates. Forces are reported for a gap $L_z=10$ µm and a wall velocity of 10 µm/s, and scale linearly with $\omega^\mathrm{w}$.}
\label{fig:fig6}
\end{figure}

We now address the disk displacement during twist-wall sweeping by quantifying the backflow generated by a translating twist wall in a planar cell. Backflow refers to flow induced by time-dependent director reorientation, even with fixed boundaries \cite{BrochardBENL,KosFGNM,BlancDONL,TothHTDN}.

\subsubsection{A sweeping twist wall as a soliton in the director field} 
We consider a twist wall separating two domains with opposite handedness and uneven twist magnitudes $\pi\pm \beta$, $\beta>0$, in a cell of thickness $L_z$. In the absence of material flow, the wall propagates because the overtwist $\lvert \beta \rvert$ produces a net elastic torque driving relaxation of the director from 
$\pi+\beta$ to $\pi-\beta$. Balancing the corresponding driving force per unit length with rotational viscous dissipation yields the wall propagation speed $\omega^\mathrm{w} = 2 \beta \pi K_{22}/\gamma L_z$ (details in SI Sec.~\ref{sec:SI_hydro}). For $K_{22}=4$ pN, $L_z=10$ µm and $\beta=10^\circ$, the propagation velocity is of the order of $5$ µm/s, a value similar to experiments (Fig. \ref{fig:fig5}B and E).

\subsubsection{Material flow is generated by the twist wall's motion} 

A moving twist wall generates material flow even in the absence of elastic stress.  We consider a steadily propagating twist wall translating with velocity $\omega^\mathrm{w}>0$, where the wall is centered at $y=0$ (see Fig. \ref{fig:fig6}A).  We assume the director structure
\begin{equation} \label{wall_ansatz}
\begin{aligned}
    n_x^{\mathrm{w}} &= \cos\!\biggl(\frac{\pi\,(z + L_z/2)}{L_z}\biggr), \\
    n_y^{\mathrm{w}} &= \sin\!\biggl(\frac{\pi\,(z + L_z/2)}{L_z}\biggr)\,\tanh\!\bigl(m\,q\,(y - \omega^{\mathrm{w}}\,t)\bigr), \\
    n_z^{\mathrm{w}} &= \varepsilon \,\sin\!\biggl(\frac{\pi\,(z + L_z/2)}{L_z}\biggr)\,\sech\!\bigl(m\,q\,(y - \omega^{\mathrm{w}}\,t)\bigr).
\end{aligned}
\end{equation}
The above ansatz, adapted from Turner, approximates the equilibrium structure of the twist wall  \cite{TurnerTWIN}, with the wall being parallel to the anchoring direction, like that in the gap above the hybrid disk when $\theta = 180^\circ$. The variables $m=\pm 1$ and $\epsilon= \pm 1$ control the handedness of each domain as well as the escaping direction (upward or downward) of the director in the wall, and $1/q$ is the characteristic thickness of the wall, which we set equal to $L_z$, as justified by Turner.  This ansatz enforces the wall's translation towards $+y$ without overtwist ($\beta =0$).

The Ericksen and Leslie formalism is adopted to model the hydrodynamics of the propagating twist wall. 
At low Reynolds number for this prescribed  director field, the model simplifies to 
\begin{align}
   \partial_j \sigma^\text{TOT}_{ij} &= 0, \label{momentum_conservation}\\
   \partial_iv_i&=0,  \label{incompressibility}\\
   n_i &= n^\mathrm{w}_i
\end{align}

The total stress is decomposed as $\sigma^{\mathrm{TOT}}_{ij}=\sigma^{\mathrm V}_{ij}+\sigma^{\mathrm E}_{ij}$. To isolate viscous backflow, we neglect elastic stresses by setting the elastic constants to zero, so that $\sigma^{\mathrm E}_{ij}=-p\,\delta_{ij}$ reduces to an isotropic pressure.
The viscous stress tensor is $\sigma^V_{ij} 
= \alpha_{1}n_{i}n_{j}n_{k}n_{l}D_{kl}
+ \alpha_{2}N_{i}n_{j}
+ \alpha_{3}N_{j}n_{i}
+ \alpha_{4}D_{ij}
+ \alpha_{5}n_{i}n_{k}D_{kj}
+ \alpha_{6}n_{j}n_{k}D_{ki}$. The term proportional to $\alpha_4 = 2\eta$ is the standard, isotropic Newtonian viscous stress, the other terms account for the anisotropic viscosity of the NLC. The vector field $\mathbf{N}$ is the co-rotational derivative of $\mathbf{n}$ measuring the rotation of $\mathbf{n}$ with respect to the vorticity of the background fluid, 
\begin{equation}
N_i = \partial_t n_i + v_j \partial_j n_i - \Omega_{ij} n_j.
\end{equation}
As shown in Fig.~\ref{fig:fig6}B and captured by our ansatz, wall propagation involves a CW rotation of the director about the $x$-axis, this temporal transformation yields a finite co-rotational derivative $\mathbf{N}$. This generates shear stresses and hence backflow, primarily through the $\alpha_2$ and $\alpha_3$ terms.
As the Stokes equations Eqs. \ref{momentum_conservation} and \ref{incompressibility} are linear in the velocity field, the solution scales linearly with the propagation speed $\omega^\mathrm{w}$, which is why we report velocity fields in terms of $\mathbf{v}/\omega^\mathrm{w}$.

In Fig. \ref{fig:fig6}C, we show the flow field solution for a twist wall separating a $\pi$-CCW twist domain for $y<0$ from a $\pi$-CW twist domain for $y>0$ ($m=+1$),  with the director escaping upward ($\epsilon =-1)$; the director field is shown in Fig. \ref{fig:fig6}B.
We find that the translation of the wall generates a localized vortex centered on the wall and decaying away from it, with peak speed $\approx0.15\omega^\mathrm{w}$. The handedness of the vortex depends on the wall topology through the sign of $m\epsilon\omega^\text{w}$ (see SI Fig. \ref{SI:hydrodynamics} A and B).

\subsubsection{Surface force on the bounding plates}
The shear flow generated by the sweeping defect results in tangential stress on the top ($f_y=-\sigma_{yz}$) and bottom plates ($f_y=+\sigma_{yz}$). In the case of infinite anchoring strength, the director field is parallel to $\mathbf{e_x}$ at both plates, and with the no-slip boundary condition, the surface stress is simply the Newtonian viscous stress associated with tangential shear flow, $\frac{1}{2}\alpha_4 \partial_z v_y$. The profiles of the surface force densities on the top and bottom plates are shown in Fig.~\ref{fig:fig6}D for $\omega^\mathrm{w}=10~\mu\mathrm{m\,s^{-1}}$. We observe that the plate towards which the director escapes is subject to a net viscous force that co-aligns with the direction of the wall motion, while the other plate feels a force in the opposite direction. In SI Sec.~\ref{sec:SI_hydro}, we consider the case of weak anchoring on the bottom plate, under which other $\alpha_i$ terms of the stress tensor are non-zero at the boundary. We find that the net force per unit length applied to the top and bottom substrates change in magnitude without changing sign (SI Fig. \ref{SI:hydrodynamics}). In the infinite anchoring limit, the net force on the bottom plate per unit length of the wall is approximately $500~\mathrm{nN\,m^{-1}}$ for $\omega^\mathrm{w}=10~\mu\mathrm{m\,s^{-1}}$.

In summary, we studied the viscous backflow generated by the translation of a mirror-symmetric twist wall in a cell with fixed boundaries. Wall motion produces a local flow vortex and shear flow that exert net viscous forces on the bounding plates. The top and bottom substrates experience opposite viscous forces, with the sign set by the topology of the wall, namely by the direction in which the director escapes across the wall. Although this calculation considers an idealized wall in a uniform cell, its magnitude is consistent with the disk motion observed experimentally. Extrapolating the force per unit length obtained for the bottom plate over the diameter of a $60~\mu\mathrm{m}$ disk gives a force scale $F_\mathrm{D,wall}\sim30~\mathrm{pN}$. Using the translational drag estimate $C_{\mathrm{D,T}}\simeq4\times10^{-5}~\mathrm{kg\,s^{-1}}$ derived in SI Sec.~\ref{sec:SI_hydro}, this force corresponds to a disk velocity $v_\mathrm{D,wall}\sim1~\mu\mathrm{m\,s^{-1}}$. This is comparable to the elastic-relaxation experiment, where $v_\mathrm{D,exp}\simeq0.5~\mu\mathrm{m\,s^{-1}}$ while the wall sweeps at $\omega^\mathrm{w}_\mathrm{exp}\simeq6~\mu\mathrm{m\,s^{-1}}$, giving the same order-of-magnitude ratio $v_\mathrm{D}/\omega^\mathrm{w}\sim0.1$.
In the experiments, the disks move opposite to the wall motion, consistent with the sign of the viscous force expected for a wall escaping from the disk surface toward the glass slide. These results show that topological instabilities, such as twist-wall motion, can generate microscopic flows in NLC devices and propel embedded objects.

\section{\label{sec:level3}Conclusion: The Defect as a Topological Flagellum}

Rotating disk colloids with hybrid anchoring carry companion topological defects that enhance translation. Together, the disk and defect periodically store and release twist elastic energy in the gap through a directed, nonreciprocal cycle of elongation, sweeping, and contraction. The sweeping is a topological instability where the defect, a non-singular twist wall, propagates as a soliton while reversing the twist handedness. These defect dynamics enhance propulsion in two ways: the sweeping wall transmits a net viscous shear to the disk through its anisotropic backflow, and the subsequent contraction sustains rotation even after the external field is removed.

These results can be viewed in the broader framework introduced by Rajabi et al. \cite{RajabiSTFS}, who showed that propulsion in nematic liquid crystals can arise when axisymmetric forcing is coupled to a surrounding orientational field with broken symmetries. Our work extends this idea by showing that topological defects can themselves drive flow and perform useful work. In this sense, the sweeping defect acts as a topological flagellum or cilium, producing thrust through the anisotropic viscous response of the nematic fluid. 

Several open questions follow from this picture. Foremost among them is the role of overtwist. In our linear hydrodynamic treatment the force the wall exerts on the disk, and hence the disk speed owing to defect sweeping, scale as the overtwist $\beta$ while the sweep duration scales as $1/\beta$, so the net per-stroke displacement is independent of $\beta$. Whether overtwist nonetheless enhances propulsion — through nonlinear hydrodynamic effects, the elastic stresses neglected in our wall ansatz — remains a candidate design parameter for future work. The same mechanism should extend to any propagating topological defect that couples to material flow — singular disclination lines as well as non-singular textures such as domain walls, kinks, and skyrmion lines.

Although our system is specific — 5CB nematic, a magnetic SU-8 disk with hybrid anchoring, and an in-plane rotating field — the underlying mechanism is not. The instability requires only that twist accumulates in a confined gap because a boundary, or the easy axis on a boundary, rotates relative to the surrounding director, and that the resulting wall couples to flow through a viscous stress. Three settings beyond our own meet these conditions. Passive lyotropic nematics sit natively in the low-$K_{22}$ regime that favors the non-singular pathway we report \cite{DietrichESTE, ZhangSATD}. Three-dimensional active nematics nucleate twist disclination lines driven by internal stresses that rotate the local director \cite{DuclosTSAD, BinyshTDAD}, and analogous wall structures are expected in polar active matter. In polar liquid-crystalline fluids, twist couples directly to a vector order parameter: ferroelectric nematics form spontaneously twisted polarization domains separated by walls \cite{LavrentovitchTSAU, SavchenkoPDTS}, while ferromagnetic nematics develop twist in response to in-plane magnetic fields, with wall motion already shown to advect embedded colloids \cite{MerteljFISO, KunduQROF}.

These findings point to strategies for soft, reconfigurable micromachines powered by structured-fluid physics, using defects introduced and controlled by optical fields, external forcing, or moving boundaries. Finally, biology suggests that collective, asynchronous defect dynamics — analogous to metachronal waves in ciliary arrays \cite{ElgetiEOMW} — could provide routes to enhanced pumping and propulsion through spatially and temporally patterned topological instabilities.

\section{\label{sec:methods}Materials and methods}

\subsection*{Fabrication of homeotropic and hybrid disk colloids}

Ferromagnetic SU-8 disk colloids of diameter $75~\mu\mathrm{m}$ and height
$30~\mu\mathrm{m}$ were fabricated by standard photolithography on silicon wafers and coated with a $\sim 15$ nm nickel layer using a Lesker PVD75 DC/RF sputterer.
For homeotropic disks, a sacrificial aluminum layer was first sputtered onto the silicon wafer before SU-8 lithography. After disk fabrication, the exposed aluminum between the disks was removed using Aluminum Etchant Type A (Transene Company, Inc.), leaving an aluminum coating on the bottom face of each disk. The disks were then coated with nickel, released from the wafer, and immersed in a 3 wt\% aqueous solution of N-dimethyl-n-octadecyl-3-aminopropyl-trimethoxy silyl chloride
(DMOAP, Sigma-Aldrich), which imposed homeotropic anchoring on all disk surfaces. The disks were finally rinsed with deionized water, dried in a vacuum oven, and dispersed in 5CB nematic liquid crystal (Sigma-Aldrich).

For hybrid disks, SU-8 disks were fabricated directly on silicon, without the sacrificial aluminum layer. After nickel coating, the disk-bearing substrate was immersed in the 3 wt\% DMOAP solution before disk release, so that only the exposed nickel-coated surfaces acquired homeotropic anchoring. The disks were then released from the wafer; the previously protected SU-8 base retained degenerate
planar anchoring, resulting in hybrid anchoring conditions. The released disks were finally dispersed in 5CB. The fabrication procedures are shown in SI Fig. S1.

\subsection*{Assembly of NLC uniform planar cells}

Glass slides were spin-coated with polyimide (PI-2555, HD Microsystems) and rubbed with a velvet cloth to impose uniform planar anchoring. Two rubbed slides were assembled in an antiparallel configuration, separated by spacers defining a $50$--$60~\mu\mathrm{m}$ gap, and glued together with ultraviolet-sensitive epoxy. The antiparallel arrangement minimized artifacts due to pretilt, thereby reducing spurious splay and bend deformations. A suspension of homeotropic or hybrid disks in isotropic 5CB was then introduced into the cell gap by capillarity.

\subsection*{Manipulated rotation of disk colloids}

Magnetic disk colloids were rotated by exposing the assembled NLC cell to a rotating magnetic field generated by a custom-built magnetic control system. The system consisted of two orthogonal pairs of electromagnetic coils (APW Company) mounted on an aluminum support around the experimental workspace. Each pair of coils was independently powered by a programmable power supply (XG 850W, Sorensen), regulated through a data acquisition board (USB-3104, Measurement Computing), and controlled
with an in-house Python script. Sinusoidal voltages of identical amplitude and relative phase shift $\pi/2$ were applied to the two coil pairs, generating a circularly rotating magnetic field with programmable stopping angle.

Imaging was performed with a high-resolution color camera (Zeiss Axiocam 705 color) mounted on an inverted microscope (ZEISS AxioVert.A1, Zeiss). The external magnetic field had magnitude $|\mathbf B|=10$ mT, well below the onset field strengths $\sim 200$--$500$ mT required for Fredericks transitions in 5CB \cite{KedzierskiOMFD,EttingerMFDD}. Consistently, the optical texture under crossed polarizers was unchanged by a steady applied field. The coil geometry was designed to produce a nearly uniform field at the center of the workspace. When a single pair of coils was energized continuously, disk drift velocities of $0.03~\mu\mathrm{m\,s^{-1}}$ were measured, providing an upper estimate of the effect of unintended field gradients. During experiments, the coils were driven sinusoidally in pairs, so the direction of any residual field gradient varied in time, further reducing its net effect.

\subsection*{Image and experimental data analysis}

The disk orientation was tracked using two visible surface markers, A and B. Defining $\mathbf r_{AB}(t)=\mathbf r_B(t)-\mathbf r_A(t)$, the disk rotation angle relative to $t=0$ was computed as $\theta(t)=\arg[\mathbf r_{AB}(t)]-\arg[\mathbf r_{AB}(0)]$, with $\arg(\mathbf r)=\operatorname{atan2}(r_y,r_x)$. Disk trajectories were tracked in ImageJ using the \textbf{Analyze Particles} function on thresholded images. For each detected disk, the centroid was recorded as the mean position of its constituent pixels, yielding time-resolved colloid positions.

\subsection*{Q-tensor numerical simulations}

The governing nonlinear partial differential equations were implemented in dimensional form and solved in COMSOL Multiphysics (Mathematics Module) using the finite-element method. The simulation box had dimensions $L_x=L_y=2~\mu\mathrm{m}$ and $L_z=1~\mu\mathrm{m}$. The top and bottom surfaces imposed planar oriented anchoring along $\mathbf e_x$ with strength $W$, whereas the four lateral sides had zero anchoring energy to model an open system. The simulated disk, with radius $R=500$ nm and height $H=500$ nm, was placed at the center of the domain, and its two circular edges were rounded with radius $R_e=50$ nm. Except for the top rounded edge, which had no anchoring, the disk surface had anchoring strength $W=10^{-3}~\mathrm{J\,m^{-2}}$. The top circular face imposed planar oriented anchoring along $\mathbf e_\theta=(\cos\theta,\sin\theta,0)$, whereas the cylindrical side wall and bottom face imposed homeotropic anchoring.

\subsection*{Flow simulation}

Steady Stokes equations for a defect wall centered at $y=0$ were solved using the Laminar Flow module of COMSOL Multiphysics. The simulation box had dimensions $L_z=10~\mu\mathrm{m}$ and $L_y=2$ mm, with no-slip boundary conditions on the bottom and top plates at $z=\pm L_z/2$ and free inlet/outlet boundary conditions at $y=\pm L_y/2$. For the Leslie viscosity coefficients $\alpha_i$, we used the values reported for 5CB by Blinov and Chigrinov \cite{BlinovEEIL,KlemanSMPA}: $\alpha_1=-0.01$, $\alpha_2=-0.08$, $\alpha_3=-0.002$, $\alpha_4=0.07$, $\alpha_5=0.1$, and $\alpha_6=-0.03$, all in $\mathrm{Pa\,s}$.

\section*{Author contributions}
Q.X.Z. performed the experiments and analyzed the data. C.D. performed the simulations, developed the theory and helped with experimental data analysis. M.R. helped with experiments and experimental data analysis. E.B.S. helped with designing the experiment. K.J.S. conceived the research and led the project. Q.X.Z., C.D. and K.J.S. wrote the manuscript. All co-authors participated in scientific discussion.

\begin{acknowledgments}
Work by C.D., M.R., Q.X.Z., and K.J.S. was supported by the U.S. Department of Energy
(DOE), Office of Science, Basic Energy Sciences (BES) under Award DE-SC0022892.
\end{acknowledgments}


\bibliography{refs}

\clearpage
\onecolumngrid

\begin{center}
{\Large \textbf{Supplementary Information}}\\[1em]
\end{center}

\setcounter{section}{0}
\renewcommand{\thesection}{S\arabic{section}}

\setcounter{figure}{0}
\renewcommand{\thefigure}{S\arabic{figure}}

\setcounter{table}{0}
\renewcommand{\thetable}{S\arabic{table}}

\setcounter{equation}{0}
\renewcommand{\theequation}{S\arabic{equation}}

\section{Testing anchoring memory on SU-8 microposts}
\label{sec:SI_anchoring}
Arrays of fixed SU-8 microposts are used to isolate the director response expected in the two gaps of fluid between a hybrid disk and the bounding surfaces. In the experiments in the main text, the hybrid disk has an untreated SU-8 face, which can acquire an oriented planar anchoring direction, and a homeotropically treated face, which produces a dipolar splay-bend distortion under a planar far-field director. Because the two faces are coupled in the rotating disk geometry, we study them separately here using post arrays with controlled surface treatments. The posts are kept fixed while the uniformly rubbed polyimide lid is rotated, allowing the relative orientation between each post surface and the imposed planar director to be varied in a controlled way. After each rotation, the system is allowed to relax before imaging. These experiments reveal the anchoring memory, twist formation, and defect rearrangements associated with each surface treatment independently of the full hybrid-disk dynamics.

\subsubsection{Untreated degenerate planar SU-8 posts}

Untreated SU-8 is expected to provide planar-degenerate anchoring. Here, we test whether cooling the nematic from the isotropic phase which is in contact with a uniformly rubbed polyimide lid can break this degeneracy and imprint a persistent in-plane anchoring direction on the SU-8 posts. Using the standard photolithographic procedure described in \hyperref[sec:methods]{Materials and Methods}, $30$ $\mu$m high cylindrical SU-8 microposts are fabricated on a glass substrate, which forms the base of the nematic liquid crystal (NLC) cell. The top substrate is a glass slide coated with uniformly rubbed polyimide (PI-2555), which imposes strong oriented planar anchoring. The top substrate is not permanently fixed and can be rotated relative to the post array.

5CB is introduced into the cell in the isotropic state, quenched into the nematic phase, and observed under crossed polarizers. In the initial configuration, denoted $0^\circ$, the rubbing direction of the top lid is parallel to one of the polarizer axes and is indicated by the arrow in Fig.~\ref{SI:rotate90andback}A. The regions above the posts appear uniformly dark, consistent with an untwisted director field aligned with the imposed rubbing direction. At this stage, the optical texture alone does not distinguish between two possibilities: the untreated SU-8 surface may remain planar-degenerate, with the local azimuthal orientation adapting to the surrounding director field, or the isotropic-to-nematic quench may have selected and retained a specific in-plane anchoring direction at the SU-8 surface. To distinguish these cases, we rotate the rubbed polyimide lid while keeping the sample in the nematic phase.

The top substrate is then rotated by $90^\circ$ while the sample remains in the nematic phase. If the SU-8 anchoring were purely planar-degenerate, the azimuthal director at the post surface could rotate with the surrounding director field, and no persistent twist would be expected above the posts. Instead, the regions above the SU-8 posts become optically bright, as shown in Fig.~\ref{SI:rotate90andback}B. This bright texture indicates the formation of twist between the rotated top lid and an in-plane anchoring direction retained at the SU-8 surface. Thus, the rotation experiment shows that the untreated SU-8 posts have acquired an oriented planar anchoring direction during the initial quench.

This imprinted anchoring direction on the SU-8 is erased by heating the sample into the isotropic phase and can be rewritten upon re-quenching under a new lid orientation, as shown in Figs.~\ref{SI:rotate90andquench}B and C. To demonstrate this, the lid is first rotated by $90^\circ$ in the nematic phase, producing the bright texture associated with twist. The cell is then heated into the isotropic phase and subsequently quenched back into the nematic phase without further rotation of the lid. After re-quenching, the previously bright regions above the SU-8 posts become dark, indicating that the twist has been eliminated and that the posts have adopted a new anchoring direction corresponding to the current orientation of the lid. Interestingly, the regions above the glass are still bright after melting and quenching, indicating that twist remains and that the untreated glass substrate has retained the initial anchoring direction.

Following rotation of the lid in the nematic phase, thin transient defect lines emerge throughout the system. These defects often emanate from the posts and can span neighboring posts. Over time, the lines shrink and annihilate as the director field relaxes. Together, these observations show that untreated SU-8 posts can be imprinted with an oriented planar anchoring state by quenching under a rubbed PI lid. This imprinted anchoring persists under mechanical rotation of the lid in the nematic phase, but is erased and rewritten by heating through the isotropic phase.

\subsubsection{Homeotropically treated SU-8 posts}

We next examine an array of homeotropically treated SU-8 microposts under a uniformly rubbed, freely rotatable polyimide lid. This experiment is designed to clarify the response expected near the homeotropic face of a hybrid disk. In particular, it tests whether the defect structure associated with a homeotropic SU-8 surface is tied to the solid object or instead selected by the far-field planar alignment imposed by the cell.

Cylindrical SU-8 posts are fabricated on a glass substrate using the procedure described above. The posts are then coated with a thin silica layer by chemical vapor deposition using silicon tetrachloride (Sigma-Aldrich), followed by treatment with DMOAP to induce homeotropic anchoring. The cell is assembled with a freely rotatable top substrate coated with uniformly rubbed PI-2555, which imposes oriented planar anchoring.

The combination of homeotropic anchoring on the posts and oriented planar anchoring on the top substrate produces a splay-bend distortion around each post. In our experiments, this distortion predominantly adopts a dipolar configuration around each post. The dipole axis is selected by the rubbing direction imposed by the polyimide lid. In the initial configuration, denoted $0^\circ$, the dipoles align with the rubbing direction and the regions above the posts appear dark under crossed polarizers, as shown in Fig.~\ref{SI:rotatewithhomeotropic}A.

The top substrate is then rotated in the nematic phase through successive orientations from $0^\circ$ to $360^\circ$. After relaxation at each orientation, the dipolar defect configurations reorient with the rubbing direction of the top lid. For example, after a $90^\circ$ rotation of the lid, the dipoles rotate by approximately $90^\circ$ in the same direction as the imposed planar alignment. This behavior is observed for lid orientations of $0^\circ$, $90^\circ$, $180^\circ$, $270^\circ$, and $360^\circ$, as shown in Figs.~\ref{SI:rotatewithhomeotropic}A--E. The regions above the posts remain dark after relaxation, indicating that no optically significant twist is retained above the homeotropic posts. Instead, the response is dominated by reorientation of the dipolar splay-bend distortion selected by the current rubbing direction of the lid.

This result indicates that the dipolar defect associated with a homeotropic SU-8 surface is not fixed with respect to the homeotropic post. Rather, its orientation is selected by the surrounding planar director field. Therefore, in the hybrid-disk experiments in the main text, where the planar cell rubbing direction is fixed while the disk rotates, the dipolar distortion below the homeotropic face is expected to remain aligned with the fixed rubbing direction of the cell rather than co-rotating with the disk. This behavior explains why the homeotropic face of the hybrid disk does not generate the same twist-memory effect as the planar face.

Rotation of the top substrate also generates thin transient defect lines. These lines separate regions in which the dipolar distortions around the posts have different orientations. For example, on opposite sides of such a line, the dipoles may point to opposite sides of the posts. As the line contracts and sweeps through the array, the local dipole orientation switches, allowing the director field to relax toward the splay/bend-dominated configuration selected by the rotated lid. This defect-mediated reorientation also provides a useful analogue for the dynamics of fully homeotropic disks rotated in a fixed planar cell: their response is governed primarily by rearrangement of the dipolar splay-bend distortion, rather than by accumulation of a persistent twist.

\section{Landau-de Gennes Q-tensor simulations}
\label{sec:Landau_de_Gennes}

To simulate director field dynamics in the absence of material flow, we rely on Landau-de Gennes theory. This framework is used to identify companion defect configurations around the disk, reconstruct their evolution under imposed disk rotation, and compute the associated free-energy changes. Nematic order is captured by the second-rank 3x3 symmetric and traceless tensor $\mathbf{Q}$, representing the second-moment of the molecular orientations

\begin{gather}
    \mathbf{Q}= \begin{pmatrix}
q_1 & q_2 & q_3\\
q_2 & q_4 & q_5\\
q_3 & q_5 & -q_1-q_4\\
\end{pmatrix}.
\end{gather}

The director field $\mathbf{n}$ and the scalar nematic order parameter $-1/2<S<1$ are retrieved by finding the leading eigenvalue of $\mathbf{Q}$ in absolute value $\lambda_1 = \frac{2}{3}S$ and the associated eigenvector of norm 1  $\mathbf{e_1}=\mathbf{n}$. In uniaxial regions, we have $ Q_{ij}=S(n_in_j-\frac{1}{3}\delta_{ij})$.

The total free energy of the NLC system in the closed domain $\Omega$ is 

\begin{equation}
    F_\text{tot}[\mathbf{Q}]=\int_{\Omega} f_{\text{bulk}}(\mathbf{Q},\nabla \mathbf{Q})\text{d}V+\int_{\partial{\Omega}}f_{\text{surf}}(\mathbf{Q})\text{d}S,
\end{equation}  

where $f_{\text{bulk}}=f_{\text{phase}}+f_{\text{elastic}}$ is the bulk free energy density. The phase free energy 

\begin{equation}
f_\text{phase}=\frac{A}{2}\mathrm{Tr}(\mathbf{Q}^2)+\frac{B}{3}\mathrm{Tr}(\mathbf{Q}^3)+\frac{C}{4}(\mathrm{Tr}(\mathbf{Q}^2))^2
\end{equation}

measures the excess of free energy with respect to the isotropic phase per unit volume. $A$ is a parameter that depends on the temperature. To model 5CB in the nematic phase, we use $A=-1.72\times 10^5 \  \mathrm{J/m^3}$, $B=-2.12\times10^6\  \mathrm{J/m^3}$, and $C=1.73\times10^6\  \mathrm{J/m^3}$, which leads to $S_0=0.8$ in the ground state in our simulations.

For the elastic free energy, we choose the 2-constant approximation

\begin{equation}
    f_\text{elastic}=\frac{L_1}{2}\partial_kQ_{ij}\partial_kQ_{ij}+\frac{L_2}{2}\partial_jQ_{ij}\partial_kQ_{ik},
\end{equation}

allowing twist and saddle-splay distortions to cost less than splay and bend distortions. The mapping with the Frank elastic constant is $K_{11}=K_{33}=S^2L_2+\frac{6S^2}{3}L_1$ and $K_{22}=K_{24}=\frac{6S^2}{3}L_1$ \cite{SussmanFSAI}. For the normal-twist penalty cost case used in Fig. 2, $L_1=6$pN and $L_2=12$pN, whereas for the low-twist penalty case in Fig. 4, $L_1=1$pN and $L_2=12$pN.

The anchoring boundary condition on a surface is enforced via the surface free energy $f_\mathrm{surf}$. To impose oriented anchoring (non-degenerate, for instance homeotropic or planar), we use the Rapini-Papoular potential

\begin{equation}
    f_\text{surf}=\frac{W}{2}\text{Tr}(\mathbf{Q}-\mathbf{Q}_\mathrm{S})^2
\end{equation}

where $\mathbf{Q}_\mathrm{S}$ is the preferred value of $\mathbf{Q}$ at the boundary, and $W$ the anchoring strength.  $\mathbf{Q}_{\mathrm{S}ij}=S_\mathrm{S}(p_ip_j-\frac{1}{3}\delta_{ij})$ is specified with the preferred director $\mathbf{p}$ and preferred order parameter $S_\mathrm{S}=0.9$ at the boundary. Since numerical simulations are typically carried at small scale, we use $W=10^{-3} \ \mathrm{J/m^2}$ to enforce strong anchoring.

The dynamics of $\mathbf{Q}$ absent material flow is governed by the gradient descent of the total free energy of the system. Each of the five independent components of $\mathbf{Q}$ ($q_i$ for $i=1,..., 5$) evolves according to

\begin{equation}
\begin{split}
    \gamma\partial_tQ_{ij}&=-[\frac{ \partial f_{\text{bulk}}}{\partial Q_{ij}}- \mathbf{\nabla}\cdot \frac{\partial f_{\text{bulk}}}{\partial \mathbf{\nabla} Q_{ij}}]  \quad  \forall \mathbf{r} \in \Omega, \\
    \text{and} \quad \gamma_\mathrm{S}\partial_tQ_{ij}&=-[\frac{\partial f_{\text{bulk}}}{\partial \mathbf{\nabla}Q_{ij}}\cdot \mathbf{\nu} +\frac{\partial f_{\text{surf}}}{\partial Q_{ij}}]    \quad \forall \mathbf{r} \in \delta\Omega
\end{split}
\label{SI:GLR}
\end{equation}

where $\mathbf{\nu}$ is the unit vector normal to the surface $\delta\Omega$ pointing outward from $\Omega$, and $\gamma$ and $\gamma_\mathrm{S}$ are the rotational viscosity coefficients in the bulk and at the surface, respectively. We use $\gamma=0.087\ \mathrm{Pa\cdot s}$ and $\gamma_\mathrm{S}=0.087\times10^{-6}\ \mathrm{Pa\cdot m\cdot s}$ so that the response time of the director to anchoring at the boundary, $\tau_\mathrm{boundary}=\gamma_\mathrm{S}/W$, is shorter than the response time associated with elastic torques in the bulk, $\tau_\mathrm{elastic}=\gamma \ell^2/L$, at the small length scales $\ell \approx 250$ nm considered here. Since the dynamics are purely relaxational, $\gamma$ and $\gamma_\mathrm{S}$ affect the relaxation time scales and transient trajectories, but not the final steady states reported here. In addition, when the evolution remains close to quasistatic, their influence is weak.\\

\textbf{Implementation in COMSOL}
The governing nonlinear partial differential equations are implemented in dimensional form and solved in COMSOL Multiphysics (Mathematics Module) using the finite-element method. The simulation box has dimensions $L_x=L_y=2~\mu\mathrm{m}$ and $L_z=1~\mu\mathrm{m}$. The top and bottom confining surfaces impose planar oriented anchoring along $\mathbf{e}_x$, while the four lateral boundaries are treated as free boundaries (zero anchoring energy) to mimic an open system. A disk of radius $R=500$ nm and height $H=500$ nm is placed at the center of the box, with its two circular edges rounded with radius $R_e=50$ nm. Except for the top rounded edge, where no anchoring is applied, the disk surface has anchoring strength $W=10^{-3}\,\mathrm{J\,m^{-2}}$: the top circular face imposes planar oriented anchoring along $\mathbf{e}_\theta=(\cos\theta,\sin\theta,0)$, while the cylindrical side wall and the bottom face impose homeotropic anchoring.

The equations are solved on a refined mesh, triangular on the boundaries and tetrahedral in the bulk. The mesh is refined near the colloid and in the gap above the disk to capture defect sweeping. Mesh refinement tests were performed to verify that the defect structure, transition angle, and free-energy evolution are not significantly altered by further mesh refinement.

Two simulation protocols are used. To obtain companion defect structures, the system is relaxed in time at fixed boundary conditions for a duration of 100 ms. To simulate driven dynamics, the anchoring direction on the top disk face is rotated continuously during a time-dependent simulation with rotation period 100 ms and total duration 120 ms. Time-dependent simulations are performed with COMSOL's implicit solver using a backward differentiation formula (BDF) time-integration scheme. The solver uses automatic ("free") time stepping, such that the internal time steps are adaptively selected during the simulation.\\

\textbf{Postprocessing}
The integrated phase, elastic, and surface free energies are computed directly in COMSOL Multiphysics. The diagonalization of $\mathbf{Q}$ to extract $S$ and $\mathbf{n}$ is also carried out in COMSOL. Director field visualizations in Figs.~1--5, vertical tilt maps, and movies are generated with COMSOL. The three-dimensional $\mathbf{Q}$-tensor field is exported from COMSOL on a regular rectangular grid and post-processed in Python to compute quantities not directly available in COMSOL, including the twist maps shown in Figs.~2 and 4. \\

\textbf{Initial Q-tensor fields used in the simulations.}
Simulations are initialized with $\mathbf{Q_0}=S_0(\mathbf{t_0}\otimes\mathbf{t_0}-\frac{1}{3}\delta_{ij})$, where the initial scalar order parameter is taken uniform $S_0=0.6$ and $\mathbf{t_0}$ is the initial director field. Because the free-energy landscape is multistable, different initial director fields $\mathbf{t_0}$ are used to converge to distinct metastable defect configurations. We detail in the next section the library of initial director fields $\mathbf{t_0}$ leading to different companion defects around disks.

\section{Companion defects of hybrid and homeotropic disks}
\label{sec:SI_companion}

We explore the library of possible director fields around hybrid and homeotropic disk colloids placed in a uniform planar cell.

The singular quadrupolar state is obtained by initializing the system with $\mathbf{t}_0=\mathbf{e}_x$. After relaxation, the hybrid disk is encircled by a half-Saturn ring at its equator. This singular disclination line, of winding number $-1/2$, is anchored to the top edge, forms two handle-like segments along the vertical sides, touches the lower edge, and then forms an additional handle beneath the homeotropic face of the disk (Fig. \ref{SI:companion_defects}A).

To access the singular dipolar state, we use the dipole ansatz 

 \begin{equation}
     \begin{split}
         t_x&=1\\
         t_y&=pR_D^2y/r^3\\
         t_z&=pR_D^2z/r^3\\
     \end{split}
 \end{equation}

from \cite{StarkPOCD}, with $p=3.08$. This singular dipolar configuration is similar to the quadrupolar state, with the half-Saturn ring displaced towards the north pole (Fig. \ref{SI:companion_defects}B). 

The coffee-cup handle up (CCHU) and down (CCHD) states for the hybrid disk are obtained from a director field defined domain-wise. It is convenient to use the cylindrical coordinates $r$, $\beta$ and $z$ defined as $r=\sqrt{x^2+y^2}$, $\beta=\mathrm{atan2}(y,x) \in [-\pi,+\pi ]$. We divide the simulation domain into subdomains and use different expressions for the initial director in each domain. In the gap above the disk, between the lid and the planar face of the disk, we set the director is parallel to $\mathbf{e_x}$:

\begin{equation}
\text{for } r < R_\text{D} \quad \text{and } z>\frac{H_\text{D}}{2} \quad  \begin{cases}
        t_x&=1\\
        t_y&=0\\
        t_z&=0 .\end{cases} 
\end{equation}   

In the gap below the disk, the director is given the splay-bend configuration:
\begin{equation}
\text{for } r < R_\text{D} \quad \text{and } z<-\frac{H_\text{D}}{2} \quad  \begin{cases}
        t_x&=\sin(\frac{\pi}{2}\frac{2z+H_D}{H_D-L_z})\\
        t_y&=0\\
        t_z&=\epsilon_{SB}\cos(\frac{\pi}{2}\frac{2z+H_D}{H_D-L_z}) \end{cases} 
\end{equation}

where $\epsilon_{SB}$ controls the direction of the splay-bend deformation. We look for a director configuration in which the topological defect is located at the north pole (near $x\approx-R_D$), so we choose $\epsilon_{SB}=+1$ so that the direction of the splay bend distortion below the disk does not necessitate a defect at the south pole of the bottom edge of the disk pole (near $x\approx+R_D$).

In the west, south and east quadrants of the disk, the director is initialized with a planar twisted configuration that goes from radial alignment in the plane $z=0$ and to uniform alignment along $\mathbf{e_x}$ at $\lvert z \lvert =\frac{L_z}{2}$
\begin{equation}
\text{for } r > R_D  \quad \text{and } \lvert \beta \lvert<\frac{3 \pi}{4} \quad  \begin{cases}
        t_x&=\cos(\beta-\beta\frac{2\lvert z\lvert}{L_z})\\
        t_y&=\sin(\beta-\beta\frac{2\lvert z\lvert}{L_z})\\
        t_z&= 0 \end{cases} 
\end{equation}

Finally, the north quadrant is initialized with two stacked twist walls running in the $x$  direction with their centers at $z=-\frac{L_z}{4}$ and $z=+\frac{L_z}{4}$:

\begin{equation}
\text{for } r > R_D  \quad \text{and } \lvert \beta \lvert>\frac{3 \pi}{4} \quad  \begin{cases}
        t_x&=\cos(\frac{2\pi z}{L_z})\\
        t_y&=\sin(\frac{2\pi z}{L_z})\tanh(qmy)\\
        t_z&=\epsilon\sin(\frac{2\pi z}{L_z})\sech(qmy)\end{cases} 
\end{equation}

To obtain the CCHU state, we set $\epsilon_{SB}=+1$, $m(z>0)=-1$ and $\epsilon(z>0)=+1$, and $m(z<0)=+1$ and $\epsilon(z<0)=-1$, so that the director escapes upward at the cores of the two twist walls stemming from the wall. After $\approx 100$ ms of relaxation, the director reaches an equilibrium configuration with no-singularity in the bulk and the two twist walls connected into a handle. Seen from profile ($x$-$z$ view) the director escapes downward in the handle, but seen from the front ($y$-$z$ view), the director escapes upward in the handle (Fig. \ref{SI:companion_defects}C). 

To obtain the CCHD state,  we set $\epsilon_{SB}=+1$,  $m(z>0)=-1$ and $\epsilon(z>0)=-1$, and $m(z<0)=+1$ and $\epsilon(z<0)=+1$, so that the director escapes downward at the cores of the two twist walls stemming from the wall. After $\approx 100$ ms of relaxation, the director reaches an equilibrium configuration with a singularity in the bulk at the north pole of the bottom edge of the disk. Seen from the front ($y$-$z$ view), the director escapes downward in the handle (Fig. \ref{SI:companion_defects}D). 

All-homeotropic disks can also support similar coffee-cup handle companion defects. For the CCHU state around the all-homeotropic disk, we use the same initial condition as the CCHU state for the hybrid disk, except that the gap above the disk is initialized with a splay-bend domain with $\epsilon_\text{SB}=-1$

\begin{equation}
\text{for } r < R_\text{D} \quad \text{and } z<-\frac{H_\text{D}}{2} \quad  \begin{cases}
        t_x&=\sin(\frac{\pi}{2}\frac{-2z+H_D}{H_D-L_z})\\
        t_y&=0\\
        t_z&=\epsilon_{SB}\cos(\frac{\pi}{2}\frac{-2z+H_D}{H_D-L_z}) \end{cases}.
\end{equation}

The CCHU configuration around the homeotropic disk has a tiny loop singularity residing at the north pole of the top edge (Fig. \ref{SI:companion_defects}E).

\paragraph{Twist around the static disk.} Unlike very thin disks and other nearly 2-D shaped objects which have been discussed in the literature, the finite thickness and homeotropic anchoring on the sides of the disks play a central role in dictating the defect structure. The homeotropic sides near the top and bottom edges of the disk require the nematogens to rotate in the gaps above and below the disks, respectively, to align with the oriented planar anchoring on the lid and base of the cell. The dipole defect originates and ends on twist walls stemming from this requirement, and the ``coffee cup handle'' structure connects these two twist walls. Consider the disk from the top; the disk's homeotropic vertical side walls generate a planar radial director field, while the base and lid impose a uniform orientation parallel to the far field director. This setting requires a twist in the director in the upper half and lower halves of the cell. There is no net twist across the cell, as the lower half and upper half are twisted with opposite handedness. There are different ways to distribute the twist around the disk for the quadrupolar and dipolar states. For the  quadrupolar state, the twist is maximum on the west and east sides, culminating at magnitude $\pi/2$, and is zero at the north and south poles (Fig. \ref{SI:companion_defects}A iv). Singular defect handles on the west and east sides separate regions with a $\pi/2$ twist of opposite handedness. For the CCHU and CCHD state, the twist increases from $0^\circ$ at the south pole to magnitude $\pi$ at the north pole, with the west side having a counterclockwise (CCW) handedness and the east side having a clockwise (CW) handedness in the upper half of the cell (Figs. \ref{SI:companion_defects}C iv and D iv). The handle dipole defect separates regions of $\pi$ twist with opposite handedness (i.e. a change of twist of  $2\pi$); the twist walls stemming from the disk edge achieve this transition via a continuous vertical tilt of the director. In the next section, we delve into the structure of twist walls and how they connect into a handle.

\section{Twist walls}
\label{sec:SI_twist_walls}

\paragraph{Single twist wall}
We study the topological structure of twist walls in a simple setting where the colloidal disk is absent. We consider a cell of thickness $h$ with $z\in[0,h]$, with planar anchoring on the top and bottom plates along the $\mathbf{e_x}$ direction. The imposed far-field domains correspond to opposite $\pi$ twists of the director about $\mathbf{e_z}$. At $y=0$, a twist wall parallel to $\mathbf{e_x}$ separates the two twist domains. The Q-tensor field is initialized with
\begin{equation}
   Q_{ij}=S_0(t_it_j-\frac{1}{3}\delta_{ij})
\end{equation}
with $S_0=0.5$, and $\mathbf{t}$ is the vector field adapted from \cite{TurnerTWIN}:
\begin{equation}
    \begin{aligned}
        t_x&=\cos(\pi z/h), \\
        t_y&=\sin(\pi z/h)\tanh(mqy), \\
        t_z&=\epsilon\sin(\pi z/h)\sech(mqy).
    \end{aligned}
    \label{SI:twist_wall_eq}
\end{equation}

The initial vector field is continuous and has unit norm everywhere; it therefore represents a nonsingular escaped twist wall rather than a disclination line. The initial vector field is continuous and has unit norm. The parameter $m=\pm1$ controls the twist handedness on either side of the plane $y=0$ and $q=\frac{1}{d}>0$ controls the thickness $d$ of the twist wall. For instance, if $m=-1$, far from the twist wall, for $y\ll -d$, the director has a planar configuration with a $\pi$-CW twist about $\mathbf{e_z}$, while in the domain $y\gg d$, the director has a planar configuration with $\pi$-CCW twist about $\mathbf{e_z}$. The transition between the two domains with opposite twist handedness occurs smoothly by rotating the director about $\mathbf{e_x}$, with a handedness controlled by $m\epsilon$, thereby tilting the director vertically in the wall. The handedness of the rotation is CCW if $m\epsilon=-1$ (the director escapes upward), and CW if $m\epsilon=+1$ (the director escapes downward). 

The simulation box has dimensions $L_x \times L_y \times h$. We use periodic boundary conditions at $x=-\frac{L_x}{2}$ and $x=+\frac{L_x}{2}$, planar oriented anchoring along the direction $\mathbf{e_x}$ at $z=0$ and $z=h$, and the boundaries at $y=-L_y/2$ and $y=+L_y/2$ impose the appropriate twisted structure:
\begin{equation}
    \begin{aligned}
        t_x&=\cos(\frac{\pi z}{h}),\\
        t_y&=\operatorname{sgn}(y)m\sin\left(\frac{\pi z}{h}\right), \\
        t_z&=0.\\ 
    \end{aligned}
\end{equation}

 We evolve the $\mathbf{Q}$-tensor field with the 2C-approximation for the elastic free energy. The final director field structure is close to the initial one. We visualize the twist wall by coloring in orange the volume in which $\lvert n_z\rvert>0.9$. Fig. \ref{SI:twist_wall} shows the twist wall structures for the four possible combinations of $m$ and $\epsilon$. 

The twist wall considered here is geometrically reminiscent of a Néel wall in ferromagnetic systems because the director rotates about the $\mathbf e_x$ axis, which lies in the wall plane. This is in contrast to a Bloch wall, where the order parameter rotates about an axis normal to the wall plane. The analogy is only structural: unlike magnetic walls, this nematic wall involves a headless three-dimensional director field and separates domains of opposite twist handedness rather than uniformly oriented domains.

\paragraph{Twist wall interactions} 
We next ask whether two vertically stacked twist-wall tubes, initially parallel to $\mathbf e_x$, can merge into a handle-like structure (Figs. \ref{SI:twist_walls_connection} and \ref{SI:twist_walls_no_connection}). To do so, we impose a patterned director field on the boundary at $x=-L_x/2$. This pattern nucleates two twist-wall tubes centered near $y=0$, one in the upper half of the cell and one in the lower half:
\begin{equation}
    \begin{aligned}
        t_x&=\cos(\frac{2\pi z}{L_z}),\\
        t_y&=\sin(\frac{2\pi z}{L_z})\tanh(m(z)qy),\\
        t_z&=\epsilon(z) \sin(\frac{2\pi z}{L_z})\sech(m(z)qy).\\ 
    \end{aligned}
\end{equation}
Here $m(z)=\pm1$ and $\epsilon(z)=\pm1$ are piecewise-constant functions that may take different values in the upper and lower halves of the cell. The function $m(z)$ sets the handedness of the far-field twist domains in each quadrant of the $(y,z)$ plane, whereas $\epsilon(z)$ sets the escape direction, or equivalently the rotation handedness, of each twist wall. The same field is used as the initial director field throughout the domain. The top and bottom plates, at $z=\pm L_z/2$, impose planar anchoring along $\mathbf e_x$. The remaining three lateral boundaries have zero anchoring strength, $W=0$. For this simulation, we used $L_x=2~\mu\mathrm m$, $L_y=4~\mu\mathrm m$, $L_z=2~\mu\mathrm m$, and $d=L_z/8$ for the thickness of the imposed twist walls.

To visualize twist walls, we color in orange the volumes where $\lvert n_z \rvert > 0.9$. NW, NE, SW, and SE refer to the quadrants of the patterned boundary in the $(y,z)$ plane when viewed from inside the simulation domain. We consider the case where $m(z>0)=-1$ and $m(z<0)=+1$. In this case, the NW quadrant of the patterned wall imposes $\pi$-CCW twist, and the SW quadrant imposes $\pi$-CW twist. Similarly, the NE quadrant of the patterned wall imposes $\pi$-CW twist, and the SE quadrant imposes $\pi$-CCW twist. Because the imposed pattern contains opposite twist domains in the upper and lower halves of the cell, it carries no net twist across the full cell thickness. The director can therefore relax toward a uniform $\mathbf e_x$ alignment far from the patterned boundary, so the twist-wall tubes stemming from the boundary must either terminate, merge, or transform as the field relaxes. The manner in which they transform depends on their relative twist wall handedness set by $\epsilon$. The clockwise/counter-clockwise handedness convention for the twist wall refers to the rotation of the director in the $(y,z)$ plane when moving along $+\mathbf{e_y}$ and viewing along $+\mathbf e_x$.
\paragraph{Case $\epsilon(z>0)=+1$ and $\epsilon(z<0)=-1$.} In this case, the two twist walls have the same handedness, \textit{i.e.}, the director rotates clockwise about the $\mathbf{e_x}$ moving along the $\mathbf{e_y}$ direction, both in the upper and lower halves of the cell. During the relaxation, we observe that the two twist walls first join to form an "X" and finally form a handle as shown in Fig. \ref{SI:twist_walls_connection}.

\paragraph{Case $\epsilon(z>0)=+1$ and $\epsilon(z<0)=+1$.} In this case, the two twist walls have opposite handedness, \textit{i.e.}, in the upper half of the cell, the director rotates clockwise from $y<0$ to $y>0$, and in the lower half of the cell, the director rotates counter-clockwise from $y<0$ to $y>0$. Initially, the upper and lower halves of the cell are mirror images of each other with respect to the $z=0$ plane. In this finite simulation domain, we do not observe a connection between the two twist-wall tubes (Fig. \ref{SI:twist_walls_no_connection}). Instead, the two structures remain separated up to the free boundary at $x=+L_x/2$. This outcome may be affected by the limited domain size and mesh resolution. In a larger or more highly resolved domain, relaxation to the uniform far field may require the formation of a localized singular region with reduced scalar order parameter.

\section{Minimal model for the disk-defect system}
\label{sec:SI_toy_model}

We denote by $\theta(t)$ the orientation of the disk relative to its zero-field equilibrium orientation. The applied magnetic field rotates in the plane with angular frequency $2\pi/T$, so that its orientation is written as $\theta_B(t)=2\pi t/T+\theta_{B0}$. The disk carries a permanent magnetic moment rigidly attached to it, whose orientation is therefore $\theta_\mu(t)=\theta(t)+\theta_{\mu0}$, where $\theta_{\mu0}$ is the angle between the magnetic moment and the reference disk orientation. We choose the initial condition where the disk is initially at its zero-field equilibrium orientation, so that $\theta(0)=0$, and such that at $t=0$, the magnetic field and the permanent magnetic moment are aligned, i.e. $\theta_B(0)=\theta_\mu(0)$, which implies $\theta_{B0}=\theta_{\mu0}$. The sweeping of the defect wall is captured via the dimensionless variable $s = \frac{S}{\pi R^2} \in [0,1]$, which is the fraction of the disk surface that has been swept by the defect.\\

We write the free energy of the system composed of the disk and the surrounding NLC as $U(s, \theta) = U_\text{magnetic}(\theta)+U_\text{elastic}(s,\theta)$. The magnetic free energy is written $U_\text{magnetic}=-\mu_B B \cos(\theta_B-\theta)$ under the agreed assumptions. The elastic free energy $U_\text{elastic}(s,\theta)=C_\text{twist}(s, \theta) + C_\text{pin}(s)$ includes the cost of twist in the gap $C_\text{twist}(s,\theta)=\frac{\pi R^2K}{2d}((1-s)\theta^2+s(\theta-2\pi)^2)$ as earlier. We consider here a a pinning energy penalizing the defect from detaching from the edge $C_\text{pin}(s) = P \sigma(\frac{s}{\epsilon})\sigma(\frac{1-s}{\epsilon})=P f(s)$, with the sigmoid function $\sigma(x)=1/(1+e^{-x})$. With $\epsilon$ small, the profile resembles a square bump with steep slope near the boundaries $s=0, 1$. The height of the energy barrier is $P$. With $\alpha = \frac{\pi R^2K}{2d}$ and $\lambda =\frac{P}{\pi\alpha}$, we have 

\begin{equation}
  U_\text{elastic}(s,\theta)=\alpha\Big( (1-s)\theta^2+s(\theta-2\pi)^2+\pi\lambda f(s)\Big).  
\end{equation}

We compare the scale of magnetic free energy to the scale of elastic free energy with $\kappa = \mu_\text{B} B / \alpha$. The full free energy of the disk-defect system is

\begin{equation}
U(s, \theta) = \alpha\Big( -\kappa \cos(\theta_B-\theta)+(1-s)\theta^2 +s(\theta-2\pi)^2+\pi\lambda f(s) \Big).
\end{equation}

We assume an overdamped dynamics, so the governing equations for the system are

\begin{align}
C_D \dot{\theta} &= -\partial_\theta U \\
C_W \dot{s} &= -\partial_s U, 
\end{align}

where $s$ is clamped clamped in the intervale $[0,1]$, ie, if $s=0$ and $-\partial_s U<0$, we enforce $\dot{s}=0$, and if $s=1$ and $-\partial_s U>0$, we force $\dot{s}=0$. $C_D$ is the  rotational drag coefficient of the disk and $C_W$ the drag opposing wall translation.

\paragraph{Elastic relaxation}
We look first into the relaxation of the disk-defect system in the elastic energy landscape in the absence of magnetic field ($B=0$) and pinning potential ($\lambda=0$). To make this set of equations non dimensional, we rescale time as $\tilde{t}= \frac{t}{\tau_\mathrm{elastic}}$, with $\tau_\mathrm{elastic}= \frac{C_D}{2\alpha}$ the elastic relaxation time of the disk. A non-dimensional parameter arise, $\mu = \frac{2 \pi C_D}{C_W}$, which compares the drag of the disk with the drag of the defect wall.  The non-dimensional set of equation is

\begin{align}
\dot{\theta} &= 2 \pi s - \theta \\
\dot{s} &= \mu(\theta-\pi)
\end{align}

We simulate several relaxation trajectories using two different initial conditions and different values for $\mu$. Trajectories are plotted over the elastic landscape in Fig. \ref{SI:elastic_relaxation}D. For a given initial condition, the relaxation pathway and the final boundary state are highly sensitive to $\mu$. This sensitivity arises because the elastic landscape contains a saddle separating the basins of attraction of the two boundary states. Changing $\mu$ changes the direction of the relaxation trajectory in the $(s,\theta)$ plane and can therefore determine on which side of the saddle the system evolves.

We estimate the rotational drag coefficient $C_D$ from the elastic relaxation time of the disk. 
Away from sweeping events, the angular relaxation is well described by
$\dot{\theta}= -\theta/\tau_{\mathrm{elastic}}$. 
Fitting this relation gives $\tau_{\mathrm{elastic}}=57~\mathrm{s}$ in the subcritical regime and 
$\tau_{\mathrm{elastic}}=62~\mathrm{s}$ in the supercritical regime 
(Fig.~\ref{SI:elastic_relaxation}C). 
Using the average relaxation time, together with the elastic energy scale 
$\alpha=4\times10^{-16}~\mathrm{J}$ estimated for a gap thickness 
$d=10~\mu\mathrm{m}$, disk radius $R=30~\mu\mathrm{m}$, and 
$K_{22}=3~\mathrm{pN}$, we obtain 
$C_D \approx 5\times10^{-14}~\mathrm{J\,s}$.

We estimate the wall drag coefficient $C_W$ from the irreversible sweeping dynamics. 
During these events, the wall sweeps at an approximately constant velocity while 
$\theta$ varies only weakly. Using averaged quantities,
\[
C_W \approx \frac{4\pi\alpha \langle \theta-\pi\rangle}{\langle \dot{s}\rangle}.
\]
With $\langle \dot{s}\rangle = 7\times10^{-2}~\mathrm{s^{-1}}$ and 
$\langle \theta-\pi\rangle = 27^\circ$, this gives
$C_W \approx 3.6\times10^{-14}~\mathrm{J\,s}$.

\paragraph{Rotation under rotating magnetic field}

When we simulate the disk under constant rotation of the magnetic-field, $\theta_B(t) = 2\pi t/T$, we render the equations non-dimensional by rescaling time as $\tilde{t}=t/T$. We identify two other timescales, the characteristic response time of the disk to the magnetic field $\tau_\mathrm{mag}=C_D/\mu_\text{B} B$ and the characteristic elastic relaxation time of the disk $\tau_\mathrm{elastic}=C_D/2\alpha$. With the 4 dimensionless parameters, $b=T/\tau_\mathrm{mag}$ and $a=T/\tau_\mathrm{elastic}$, $\mu=\frac{2 \pi C_D}{C_W}$ and $\lambda=\frac{P}{\pi \alpha}$, we obtain 

\begin{align}
\dot{\theta} &= b\sin(2\pi \tilde{t}-\theta)+a(2 \pi s - \theta),\\
\dot{s}&= \mu a (\theta-\pi)-\frac{1}{4}\mu a \lambda g(s),
\end{align}

where $g(s)=\frac{1}{\epsilon}\sigma(\frac{s}{\epsilon})\sigma(\frac{1-s}{\epsilon}) \left(\sigma(\frac{1-s}{\epsilon})-\sigma(\frac{s}{\epsilon})\right)$.
The first equation is the balance of the magnetic, elastic and viscous torque on the disk. The second equation is the balance of effective elastic and viscous forces acting on the defect.

We analyze the system's stability upon variation of the magnetic field direction $\theta_B$ and the magnetic field strength $\kappa=\frac{2b}{a}=\frac{\mu_B B}{\alpha}$. To simplify, we do not consider the pinning potential in this part ($\lambda=0$). The free energy we consider is 

\begin{equation}
U(s, \theta) = \alpha\Big( -\kappa \cos(\theta_B-\theta)+(1-s)\theta^2 +s(\theta-2\pi)^2 \Big).
\end{equation}

We first look for interior fixed points (minima, maxima, saddle points), which are points $s^\star, \ \theta^\star$ where  $\mathbf{\nabla} U =0$. This leads to $\theta^\star = \pi$ and $-\kappa \sin(\theta_B-\pi)+2(\pi-2\pi s^\star)=0$. Their stability can be determined by the sign of the determinant of the Hessian matrix 

\begin{equation}
\mathbf{H}(\theta,s)=
\begin{pmatrix}
\partial_{\theta \theta} U & \partial_{\theta s} U\\
\partial_{\theta s} U & \partial_{s s} U
\end{pmatrix}
=
\alpha
\begin{pmatrix}
\kappa \cos(\theta_B-\theta)+2 & -4\pi\\
-4\pi & 0
\end{pmatrix} 
\end{equation}.

Since $\text{det}\mathbf{H}= -(4\pi \alpha)^2 <0$, interior fixed points, if they exist, are saddle-points. 

Since $s \in [0,1]$ via clamped dynamics, we also have to consider boundary fixed points, which satisfy a less constraining condition:
\begin{itemize}
    \item for $s=0$, $\partial_\theta U = 0$ and $\partial_s U \geq 0$,
    \item for $s=1$, $\partial_\theta U = 0$ and $\partial_s U \leq 0$.
\end{itemize}

Boundary fixed points $\theta^{\star,0}$ on the $s=0$ boundary satisfy $-\kappa \sin(\theta_B-\theta^{\star,0})+2\theta^{\star,0}=0$ and $\theta \leq \pi$. Boundary fixed point $\theta^{\star,1}$ on the $s=1$ boundary satisfy $-\kappa \sin(\theta_B-\theta^{\star,1})+2(\theta^{\star,1}-2\pi)=0$ and $\theta \geq \pi$.
A boundary fixed point is a minimum (stable) if $\partial_{\theta \theta}U >0$, and a boundary saddle (unstable) if  $\partial_{\theta \theta}U <0$. On bifurcation diagrams, we plot the stable branch in orange and turquoise for the $s=0$ and $s=1$ boundary respectively, while the branches of boundary saddle-points are plotted with red dashed and dash-dotted lines. The bifurcation diagram vs the magnetic field strength $\kappa$ for the symmetric case $\theta_B=\pi$ is a pitchfork bifurcation (Fig. \ref{SI:bifurcation_diagrams} A i); when the magnetic field breaks the symmetry the pitchfork is broken (Fig. \ref{SI:bifurcation_diagrams} A ii). The bifurcation diagram vs $\theta_B$ at fixed magnetic field strength $\kappa$ is particularly useful to understand the dynamics during the driven rotation of the disk. It displays disconnected stable branch, and predicts that defect sweeping is an out-of-equilibrium switching event between competing stable boundary fixed points.

A pinning energy for the defect delays the onset of sweeping to an angle $\theta_\mathrm{unpin}>\pi$. The theoretical prediction for the sigmoid profile used here is $\theta_\text{unpin}=\pi + \frac{\lambda}{16\epsilon}$. In numerical simulations, we detect the onset of sweeping using the criterion $s>s_\mathrm{threshold}$, with $s_\mathrm{threshold}=5\times10^{-2}$. We compare the measured unpinning angle with the theoretical prediction for two values of $\epsilon$ in Fig.~\ref{SI:pinning_energy}.

\section{Hydrodynamics of a translating twist wall}
\label{sec:SI_hydro}

\paragraph{Model and assumptions}
We consider a nematic layer invariant along the $x$ axis, unbounded along $y$, and confined between a substrate and a lid at $z=-L_z/2$ and $z=L_z/2$. A twist wall, parallel to $\mathbf e_x$, separates two $\pi$-twist domains of opposite handedness. We ask what flow is generated when this wall translates. To address this question, we use a simplified Ericksen--Leslie description of an incompressible nematic, in which the director field is prescribed and the back-coupling of the flow on $\mathbf n$ is neglected. This description is appropriate here because the wall is non-singular and can be represented by a continuous director field at fixed scalar order parameter.

\paragraph{Overtwist drives twist-wall motion}
We first estimate the effective elastic driving force per unit length acting on a twist wall separating domains of unequal twist. Consider a cell of thickness $d$ in which the wall separates two twist domains: for $y>0$, a domain of net twist $\pi+\beta$, and for $y<0$, a domain of net twist $-\pi+\beta$. Their elastic energies per unit area are
\begin{equation}
f_1=\frac{K}{2}\frac{(\pi+\beta)^2}{d}, \qquad
f_2=\frac{K}{2}\frac{(-\pi+\beta)^2}{d}.
\end{equation}
If the wall is displaced by $\mathrm{d}y$, the total elastic free energy per unit length varies by
\begin{equation}
\mathrm{d}F_\mathrm{tot}= -f_1\,\mathrm{d}y + f_2\,\mathrm{d}y,
\end{equation}
so that the resulting effective force per unit length along $\mathbf e_y$ is
\begin{equation}
F_\mathrm{E}=-\frac{\mathrm{d}F_\mathrm{tot}}{\mathrm{d}y}
=\frac{2K\pi}{d}\beta.
\end{equation}
For $\beta>0$, the wall is driven toward $y>0$, i.e. in the direction that shrinks the more strongly twisted domain. For $d=10~\mu\mathrm{m}$, $K=4$ pN, and $\beta=6^\circ$, this gives $F_\mathrm{E}\sim 10^{-7}~\mathrm{N\,m^{-1}}$. Balancing this driving force with viscous drag gives the scaling estimate
\begin{equation}
\omega^\mathrm{w}\sim \frac{2\pi K \beta}{\gamma d},
\end{equation}
where $\gamma=0.087~\mathrm{Pa\,s}$ is the rotational viscosity of 5CB. Having established that overtwist drives wall propagation, we now consider a wall with symmetric twist profile ($\beta=0$) translating at prescribed speed $\omega^\mathrm{w}$ in order to isolate the associated viscous backflow.

\paragraph{Translating twist-wall ansatz}

In the case of the hybrid disk, the anchoring is weaker at the disk surface (bare SU-8) than at the top plate (rubbed polyimide). To take into account this asymmetry into our model, we consider a top plate with infinitely strong planar anchoring along $\mathbf e_x$ and a bottom plate with finite planar anchoring strength $W$ along $\mathbf e_x$.  Introducing the extrapolation length $\ell=K_{22}/W$, we prescribe the translating director field as
\begin{align}
n_x^{\mathrm{w}} &= \cos\!\biggl(\frac{\pi\,(z + \ell + L_z/2)}{L_z +\ell}\biggr), \\
n_y^{\mathrm{w}} &= \sin\!\biggl(\frac{\pi\,(z + \ell + L_z/2)}{L_z +\ell}\biggr)\,\tanh\!\bigl(m\,q\,(y - \omega^{\mathrm{w}}\,t)\bigr), \\
n_z^{\mathrm{w}} &= \varepsilon \,\sin\!\biggl(\frac{\pi\,(z + \ell + L_z/2)}{L_z + \ell}\biggr)\,\sech\!\bigl(m\,q\,(y - \omega^{\mathrm{w}}\,t)\bigr).
\end{align}
This form represents two twist domains of opposite handedness separated by a wall centered at $y^{\mathrm w}(t)=\omega^{\mathrm w} t$, with an out-of-plane escape localized near the wall. It satisfies planar alignment along $\mathbf e_x$ at $z=+L_z/2$ and at the virtual plane $z=-L_z/2-\ell$ associated with the extrapolation length. The strong-anchoring limit at the bottom plate is recovered for $\ell\to 0$.

\paragraph{Flow equations}
We want to check whether such dynamics of the director generates a flow in the cell and a net force on the plates. We impose the dynamics of the director $\mathbf{n}$ and we neglect the feedback of the flow on the director. At low Reynolds number, the equations for the flow are the Stokes equation and the incompressibility condition
\begin{align}
   \partial_j \sigma^\text{TOT}_{ij} &= 0, \\
   \partial_iv_i&=0, \\
   n_i &= n^\mathrm{w}_i
\end{align}

Our convention for the stress tensor is that $\sigma_{ij}$ is the force per unit surface exerted in the direction $\mathbf{e_i}$ on the surface element with normal $\mathbf{e_j}$, consistent with the formalism of Denniston \cite{DennistonTASO}. 
The total stress tensor contains viscous and elastic contributions. Here, to isolate viscous backflow generated by a prescribed translating wall, we neglect the Ericksen stress and write
\begin{equation}
\boldsymbol{\sigma}^\mathrm{TOT}=\boldsymbol{\sigma}^\mathrm{V}-p\mathbf I .
\end{equation}

In the Ericksen-Leslie model, the viscous stress tensor is 
\begin{equation}
\sigma^V_{ij} 
= \alpha_{1}n_{i}n_{j}n_{k}n_{l}D_{kl}
+ \alpha_{2}N_{i}n_{j}
+ \alpha_{3}N_{j}n_{i}
+ \alpha_{4}D_{ij}
+ \alpha_{5}n_{i}n_{k}D_{kj}
+ \alpha_{6}n_{j}n_{k}D_{ki},
\end{equation}

with the strain rate tensor $D_{ij}=\frac{1}{2}(\partial_iv_j+\partial_jv_i)$. We recognize the Newtonian viscous stress $\alpha_{4}D_{ij}=2\eta D_{ij}$.

This calculation isolates the viscous backflow generated by a translating wall of prescribed shape and speed; it does not describe the full coupled elastohydrodynamic selection of the wall motion.

\paragraph{Numerical implementation}
We solve the Stokes problem numerically using the laminar flow module of COMSOL Multiphysics. The computational domain has thickness $L_z=10~\mu$m and length $L_y=2$ mm. No-slip boundary conditions are imposed at the bottom and top plates, $z=\pm L_z/2$, and free inlet/outlet conditions are applied at $y=\pm L_y/2$. The Leslie viscosity coefficients are taken from measurements on 5CB reported by Blinov and Chigrinov \cite{BlinovEEIL, KlemanSMPA}: $\alpha_1=-0.01$, $\alpha_2=-0.08$, $\alpha_3=-0.002$, $\alpha_4=0.07$, $\alpha_5=0.1$, and $\alpha_6=-0.03$ Pa$\cdot$s. 

We decompose the stress tensor into a Newtonian part, $\sigma^\mathrm{N}_{ij}=-p\delta_{ij}+2\eta D_{ij}$, and a nematic contribution, $\sigma^\mathrm{NLC}_{ij}=\sigma^\mathrm{TOT}_{ij}-\sigma^\mathrm{N}_{ij}$. The bulk force density $\partial_j\sigma^\mathrm{NLC}_{ij}$, evaluated at $t=0$ for a wall centered at $y=0$ and moving with speed $\omega^\mathrm{wall}$, is then supplied to COMSOL as a volume force. The fluid is initially at rest, and the flow relaxes rapidly toward a steady state. In practice, the steady solution is obtained with a time-dependent implicit solver (BDF, fully coupled), using free time stepping over a total simulation time of $10$ ns. The velocity field converges within a few nanoseconds to a localized vortex centered on the wall.

\paragraph{Hydrodynamic flow and surface force}

Wall translation generates a localized vortex in the $yz$ plane, centered on the wall, together with shear stresses on the confining plates. The handedness of the vortex is set by the sign of $m\varepsilon\omega^\mathrm{w}$, and therefore depends on both the wall topology and its propagation direction.

Under our stress convention, the traction exerted by the fluid on a surface element with inward normal $\mathbf{dS}$ is $\boldsymbol{\sigma}^\mathrm{TOT}\mathbf{dS}$. We are interested in the tangential traction along $\mathbf e_y$ on the two bounding plates, namely
\begin{equation}
f_{y,\mathrm{bot}}(y)=\sigma_{yz}(y,z=-L_z/2), \qquad
f_{y,\mathrm{top}}(y)=-\sigma_{yz}(y,z=+L_z/2).
\end{equation}
The corresponding profiles are shown in Fig.~S16E for decreasing anchoring strength at the bottom plate. Integrating these profiles over $y$ gives the net viscous force per unit length along the invariant $x$ direction.

In the strong-anchoring limit, the director is parallel to $\mathbf e_x$ at the plates, so the tangential traction is dominated by the Newtonian shear term, $\sigma_{yz}\simeq \frac{1}{2}\alpha_4 \partial_z v_y$. The resulting viscous force scales linearly with the wall speed. For a wall propagating at $10~\mu\mathrm{m\,s^{-1}}$, we obtain a net force per unit length of order $F_\mathrm{viscous}\approx 5\times10^{-7}~\mathrm{N\,m^{-1}}$. The top and bottom plates experience viscous forces of opposite signs, consistent with the antisymmetry of the induced shear flow across the cell. Reversing the wall escape direction reverses the sign of the viscous traction.

This viscous contribution should be distinguished from the effective elastic driving force that causes the wall to propagate in the first place, and for which the resulting substrate force is expected to oppose that motion. The latter is set by the overtwist/undertwist imbalance discussed above, whereas the viscous traction reported here is the hydrodynamic response to an imposed translating wall. The direction of the viscous traction is sensitive to the wall escaping direction. In the experiment, the observed disk motion is opposite to the wall sweep, which is consistent with the case of a director escaping from the disk surface toward the glass slide in the wall.

\paragraph{Effect of finite anchoring at the bottom plate}

When the bottom anchoring is weak, the director at the plate can deviate from $\mathbf e_x$, and additional Leslie-stress terms become nonzero at the boundary. In our simulations, the net force per unit length increases in magnitude on the weak-anchoring plate and decreases in magnitude on the strong-anchoring plate, as shown in Figs.~\ref{SI:hydrodynamics}E and F. Importantly, over the range explored here, the viscous surface force does not reverse sign as the bottom anchoring is reduced.

\paragraph{Role of Ericksen stress}
In our analysis, we deliberately neglect the elastic contribution to the stress in order to isolate the viscous backflow generated by the prescribed motion of the twist wall. In a full Ericksen--Leslie description, elastic stresses enter through the Ericksen stress tensor, $\mathrm{\sigma}^E_{ij}
= -\frac{\delta F_{\mathrm{bulk}}}{\delta(\partial_j n_k)}\partial_i n_k
- p\delta_{ij}$, where $f$ is the Frank free-energy density and $p$ is the pressure enforcing incompressibility. This stress depends only on the instantaneous director configuration and, if present, on external fields through their contribution to the free energy. 

For a flat plate with uniform strong anchoring, and in the absence of external fields, the tangential Ericksen traction $\sigma^{E}_{yz}=0$ exerted directly on the plate vanishes. This result follows from the fact that the surface director does not vary along the tangential directions, so that the tangential derivatives entering $\sigma^{E}_{yz}=0$ are zero at the boundary. The free energy is locally invariant upon tangential translation of the plate, so the plate is not subject to the tangential elastic stress directly. Nonetheless, an elastic stress can be present in the bulk and be a source of material flow in the cell, which results in a viscous stress on the plates. For a twist wall with mirror symmetry with respect to $x,z$ plane ($\alpha=0$), the flow from Ericksen stress has the same symmetry, resulting in zero net force on each plate. When the wall separates domains of unequal twist, $\alpha\neq0$, this symmetry is broken. The Ericksen stress is then expected to drive a biased recirculating flow around the wall, which leads to a net viscous tangential stress on the plate. 

\paragraph{Transposing these results to the disk}
Our simulations show that a twist wall propagating at $\omega^\mathrm{w}=10~\mu\mathrm{m\,s^{-1}}$ exerts a net viscous force of backflow origin on a plate of approximately $500~\mathrm{nN\,m^{-1}}$ per unit length of wall (Fig.~\ref{SI:hydrodynamics}F). For a hybrid disk of diameter $\simeq 60~\mu\mathrm{m}$, this corresponds to a force scale $F_\mathrm{D, wall}\approx 30$ pN. We estimate the resulting disk velocity by balancing this force with the translational viscous drag on the disk. The dominant contribution to this drag is expected to arise from Couette-like shear flows in the gaps above and below the disk, each of thickness $h=10$µm. This gives $C_{\mathrm{D,T}}
\simeq
\frac{2\pi \eta R^2}{h}
\simeq
4\times 10^{-5}~\mathrm{kg\,s^{-1}}$, and therefore
$v_\mathrm{D,wall}
=
\frac{F_\mathrm{D,wall}}{C_{\mathrm{D,T}}}
\simeq
1~\mu\mathrm{m\,s^{-1}}.
$
In the elastic-relaxation experiment, the disk translates at $v_\mathrm{D,exp}\simeq0.5~\mu\mathrm{m\,s^{-1}}$ while the wall sweeps across the disk at $\omega^\mathrm{w}_\mathrm{exp}\simeq6~\mu\mathrm{m\,s^{-1}}$. Thus, both the simulation-based estimate and the experiment give a comparable ratio, $\frac{v_\mathrm{D}}{\omega^\mathrm{w}}\sim 0.1.$
This agreement suggests that viscous backflow generated by the sweeping twist wall can account for the observed order of magnitude of disk translation. In the experiment, however, this contribution is not isolated: additional forces may act simultaneously on the disk, including viscous stresses of elastic origin and hydrodynamic forces associated with the disk rotation itself.

\begin{figure}
\centering
\includegraphics[width=\textwidth]{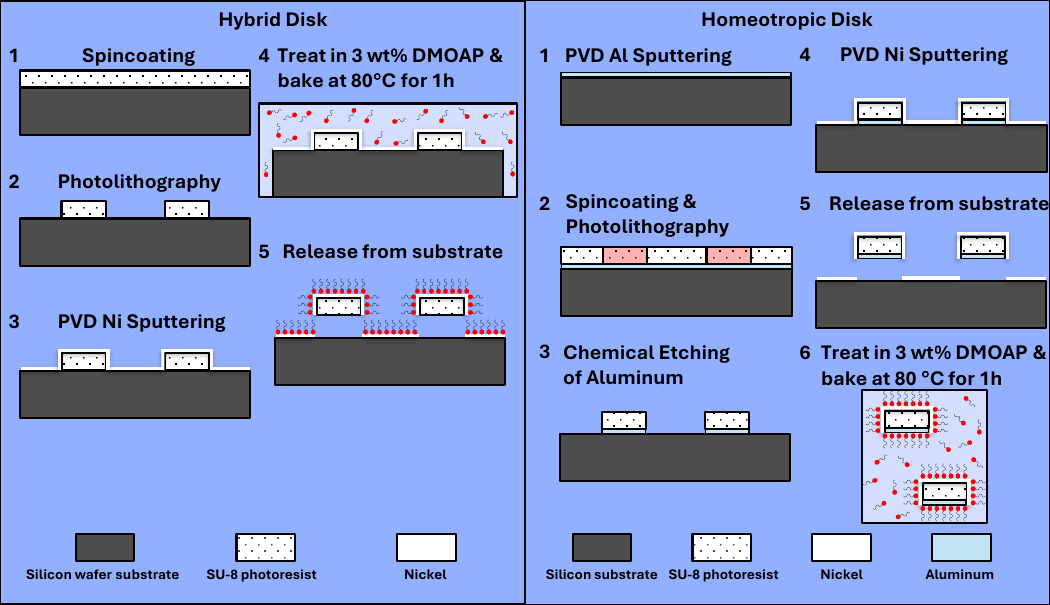} 
\caption{Detailed schematic of the fabrication process of hybrid and homeotropic disks.}
\label{SI:schematic}
\end{figure}

\begin{figure}
\centering
\includegraphics[width=\textwidth]{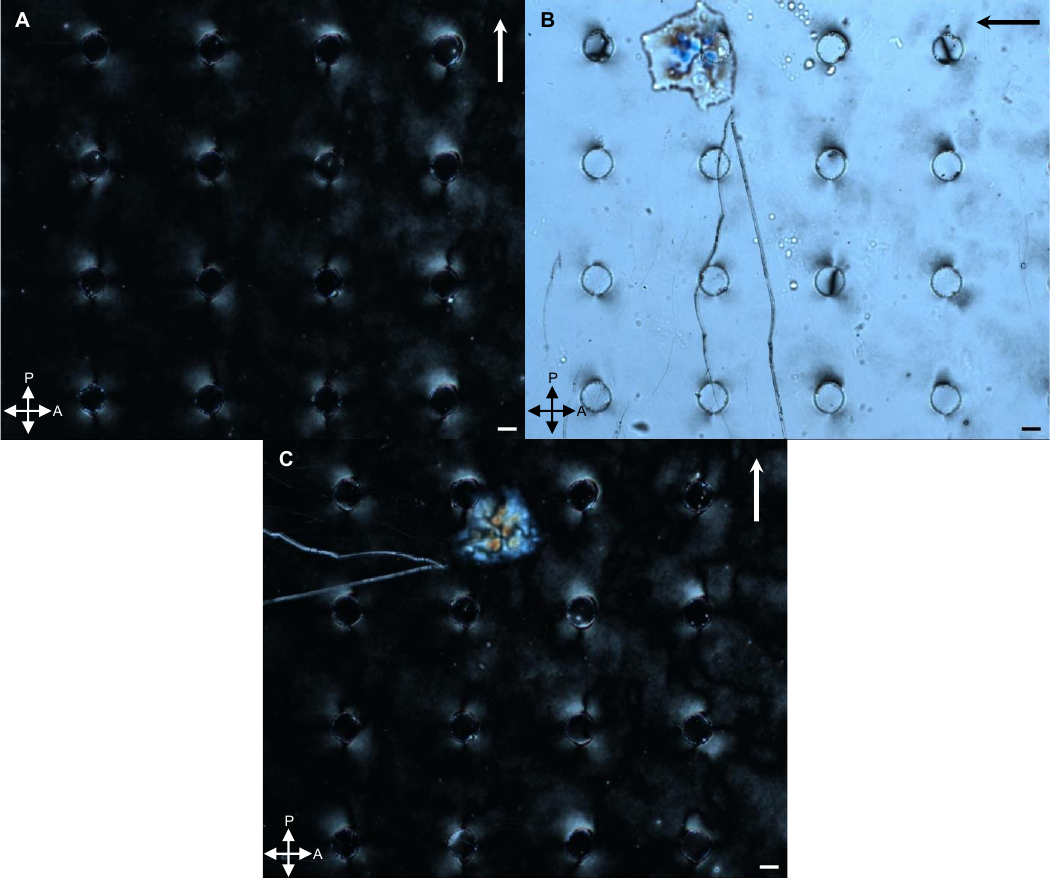}
\caption{
Mechanical rotation of the rubbed lid reveals anchoring memory on untreated SU-8 posts. Experimental snapshots of an untreated SU-8 array of cylindrical posts observed under crossed polarizers.  The polyimide-coated lid imposes uniform planar anchoring, with the rubbing direction indicated by the arrow. (A) In the initial configuration, denoted $0^\circ$, the rubbing direction is vertical and parallel to one polarizer axis, giving a dark texture above the posts. (B) When the lid is rotated by $90^\circ$ counterclockwise while the sample remains in the nematic phase, the regions above the posts become bright, indicating the formation of twist between the rotated lid and the in-plane anchoring direction retained at the SU-8 surface. (C) Rotating the lid back to $0^\circ$ removes the twist and restores the dark texture.
Scale bars, $50~\mu\mathrm{m}$.
}
\label{SI:rotate90andback}
\end{figure}

\begin{figure}
\centering
\includegraphics[width=\textwidth]{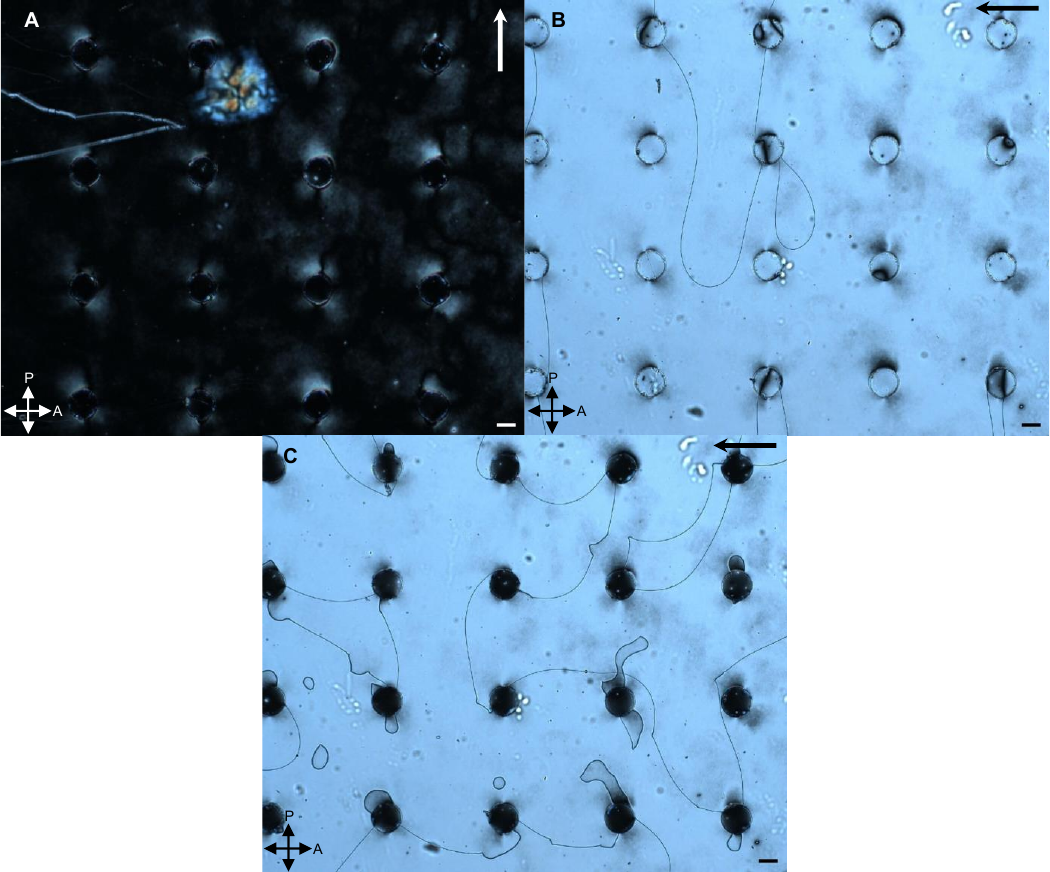}
\caption{Heating through the isotropic phase erases the anchoring direction imprinted on untreated SU-8 posts. Experimental snapshots of an untreated SU-8 post array observed under crossed polarizers. The polyimide-coated lid imposes uniform planar anchoring, with the rubbing direction indicated by the arrow. (A) In the initial $0^\circ$ configuration, the rubbing direction is vertical and the regions above the posts appear dark. (B) The lid is rotated by $90^\circ$ counterclockwise in the nematic phase, producing a bright texture above the posts that indicates twist. (C) The cell is then heated into the isotropic phase (not shown) and quenched back into the nematic phase without further rotation of the lid. After re-quenching, the regions above the posts become dark at the $90^\circ$ lid orientation, indicating that the previous twist has been erased and that the SU-8 surface has adopted a new in-plane anchoring direction selected by the current lid orientation. The region above the glass are still bright, indicating that the untreated glass substrate has retained the initial anchoring direction.
Scale bars, $50~\mu\mathrm{m}$.
}
\label{SI:rotate90andquench}
\end{figure}

\begin{figure}
\centering
\includegraphics[width=\textwidth]{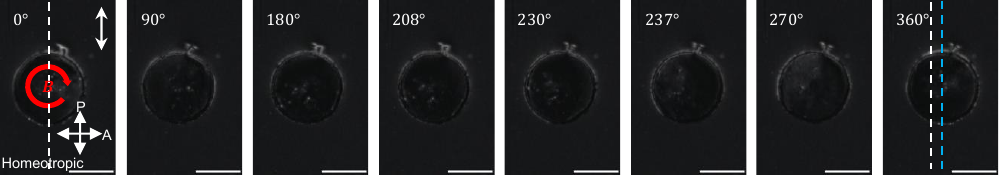}
\caption{Cross-polarization microscopy of the homeotropic disk rotating one full revolution over one period ($T=45$ s). White and blue dashed lines indicate the initial and final position of the center-of-mass (COM) of the disk, respectively.}
\label{SI:homeotropicXPOL}
\end{figure}

\begin{figure}
\centering
\includegraphics[width=\textwidth]{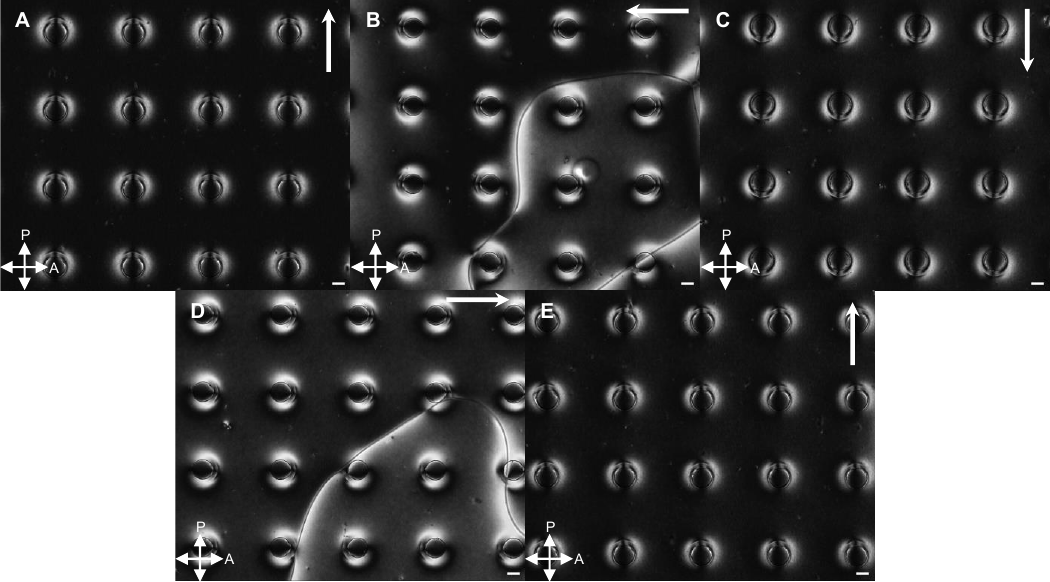}
\caption{
Homeotropically treated SU-8 posts reorient with the rubbed lid without retaining twist.
Experimental snapshots of a homeotropically treated SU-8 post array observed under crossed polarizers.
The polyimide-coated lid imposes uniform planar anchoring, with the rubbing direction indicated by the arrow.
The lid is successively oriented at (A) $0^\circ$, (B) $90^\circ$, (C) $180^\circ$, (D) $270^\circ$, and (E) $360^\circ$.
At each orientation, after relaxation, the regions above the posts remain dark, indicating that no optically significant twist is retained above the homeotropic posts.
Instead, the response is consistent with reorientation of the dipolar splay-bend distortion selected by the current rubbing direction of the lid.
This behavior provides an analogue for the homeotropic face of hybrid disks, where the dipolar distortion is expected to remain aligned with the fixed far-field rubbing direction rather than co-rotating with the solid disk.
Scale bars, $50~\mu\mathrm{m}$.
}
\label{SI:rotatewithhomeotropic}
\end{figure}

\begin{figure}
\centering
\includegraphics[width=1\textwidth]{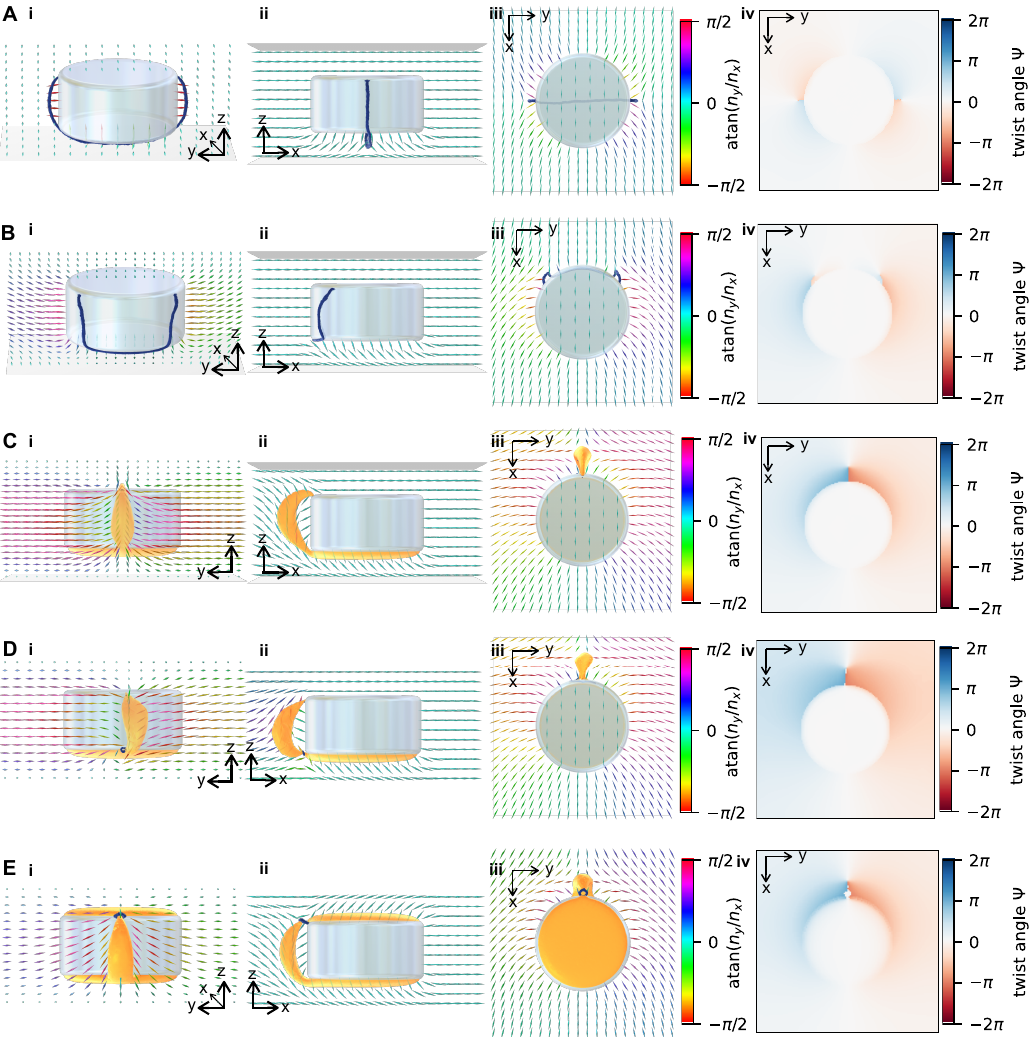}
\caption{Possible companion defects of unrotated hybrid and homeotropic disk in simulations. (A) Singular quadrupolar state of a hybrid disk. (B) Singular dipolar state of a hybrid disk. (C) Non-singular CCHU state of a hybrid disk. (D) CCHD state of a hybrid disk. (E) CCHU state of a all-homeotropic disk. For each case: i) $y-z$ view, ii) profile $x-z$ view, iii) top $x-y$ view, iv) net twist angle in the upper half of the cell ($z>0$).}
\label{SI:companion_defects}
\end{figure}

\begin{figure}
    \centering
    \includegraphics[width=1\textwidth]{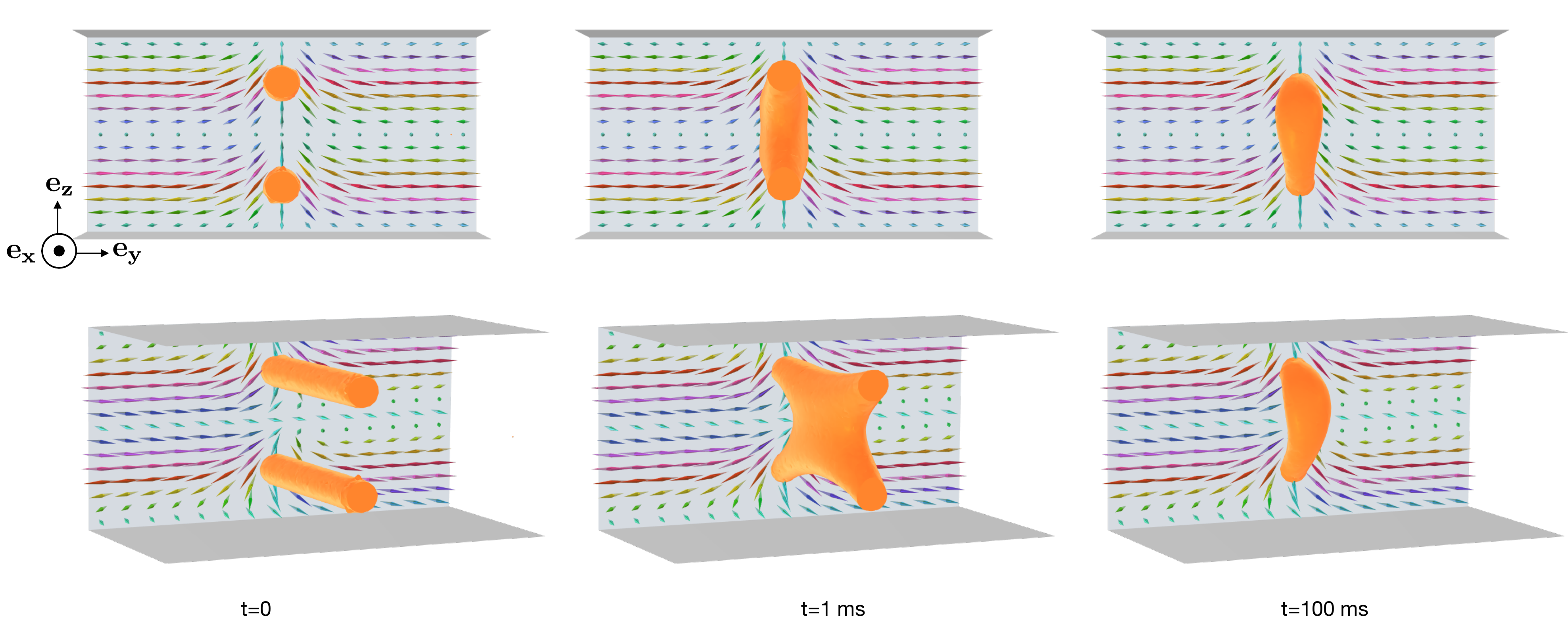}
    \caption{The handedness of co-escaping twist walls mutually agree and twist walls connect into a handle after relaxation.}
    \label{SI:twist_walls_connection}
\end{figure}

\begin{figure}
    \centering
    \includegraphics[width=1\textwidth]{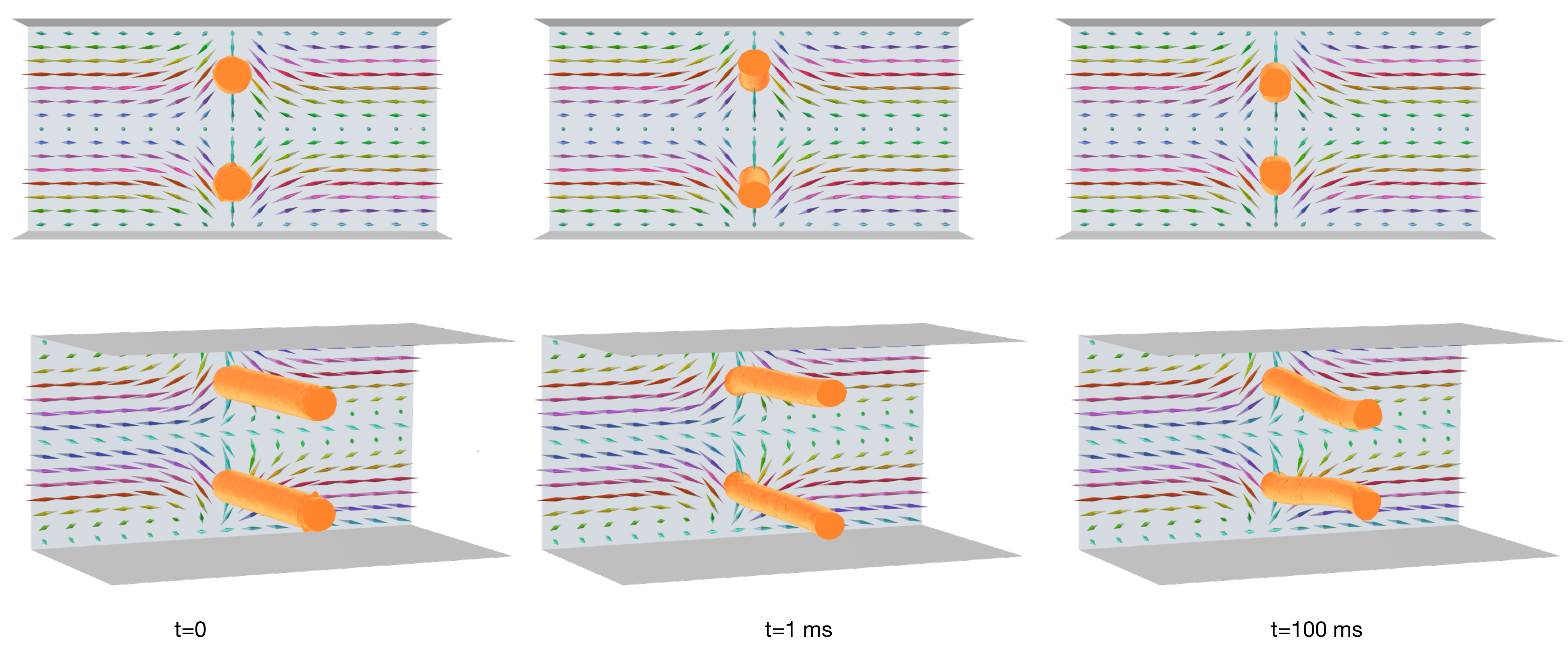}
    \caption{The handedness of anti-escaping twist walls are opposite and the twist walls do not connect to form a smooth connection.}
    \label{SI:twist_walls_no_connection}
\end{figure}

\begin{figure}
\centering
\includegraphics[width=\textwidth]{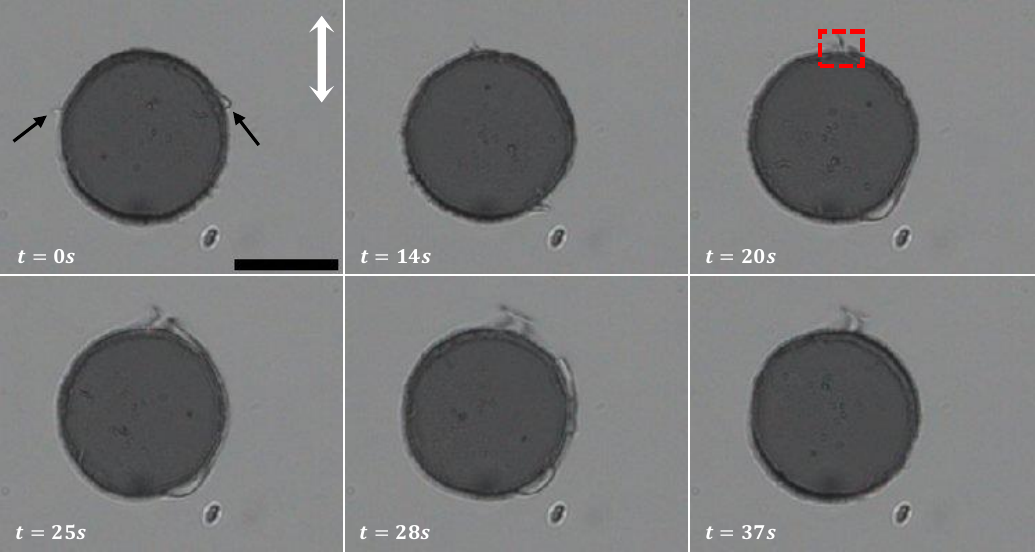}
\caption{Time-stamped sequence of a homeotropic disk initially exhibiting a quadrupolar Saturn-ring defect marked by black arrows, which subsequently transforms into a dipole defect. The red box marks the location where the dipole defect starts emerging from the Saturn-ring loop. Scale bar, $50$ µm.}
\label{SI:homeotropictransition}
\end{figure}

\begin{figure}
    \centering
    \includegraphics[width=1\textwidth]{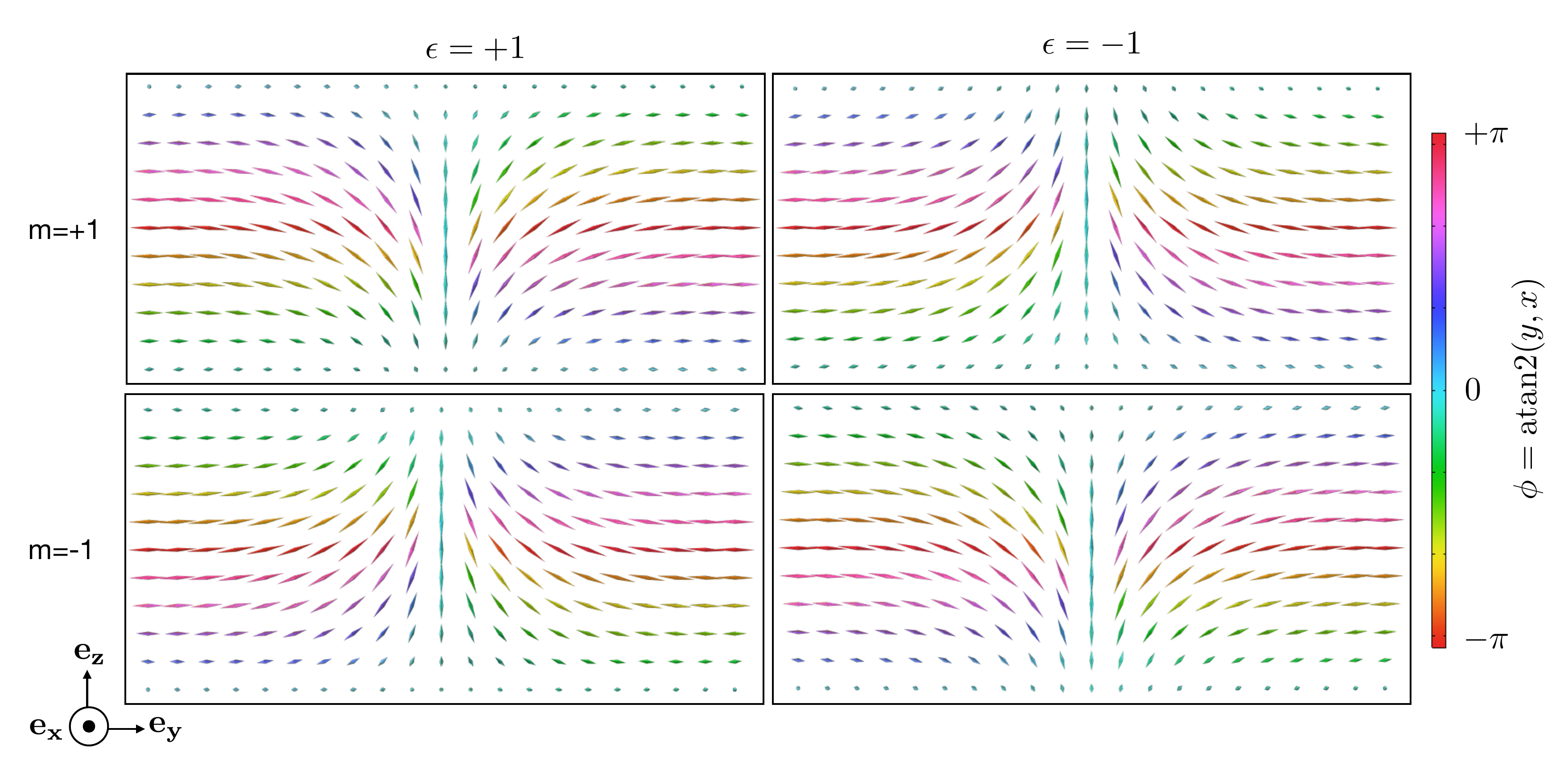}
    \caption{Structure of the twist walls for four possible pairs of $m$ and $\epsilon$, which all have similarities with Neel walls, where the transition from one domain to the other occurs by rotating the vector about the $x$-axis which is in the plane of the domain wall ($xz$-plane).}
    \label{SI:twist_wall}
\end{figure}

\begin{figure}
    \centering
    \includegraphics[width=1\textwidth]{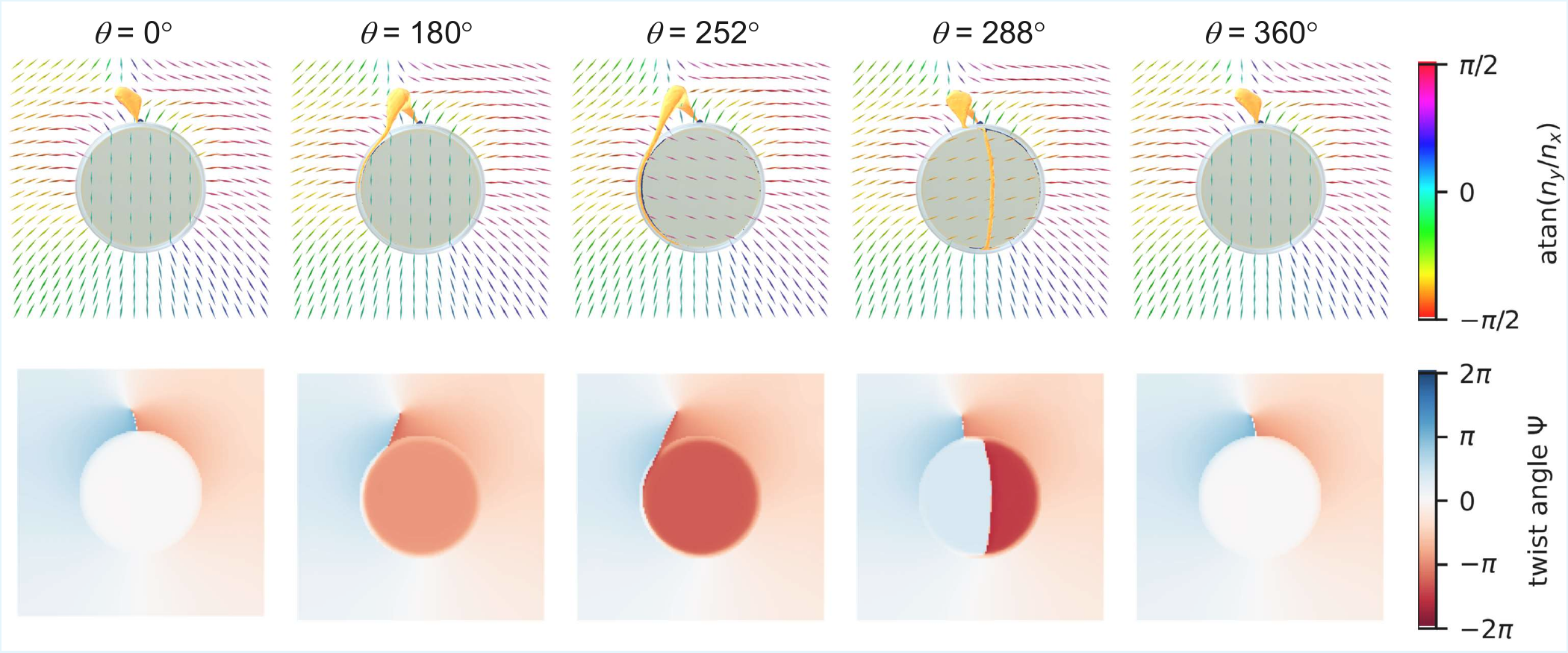}
    \caption{Director around the disk at $z=0$ and at the surface of the planar face of the disk during a full rotation and corresponding map of the total twist of the director from $z=0$ to the top lid at $z=H_\mathrm{gap}/2$ for a CCHD configuration. }
    \label{SI:CCHD_rotation}
\end{figure}

\begin{figure}
    \centering
    \includegraphics[width=1\textwidth]{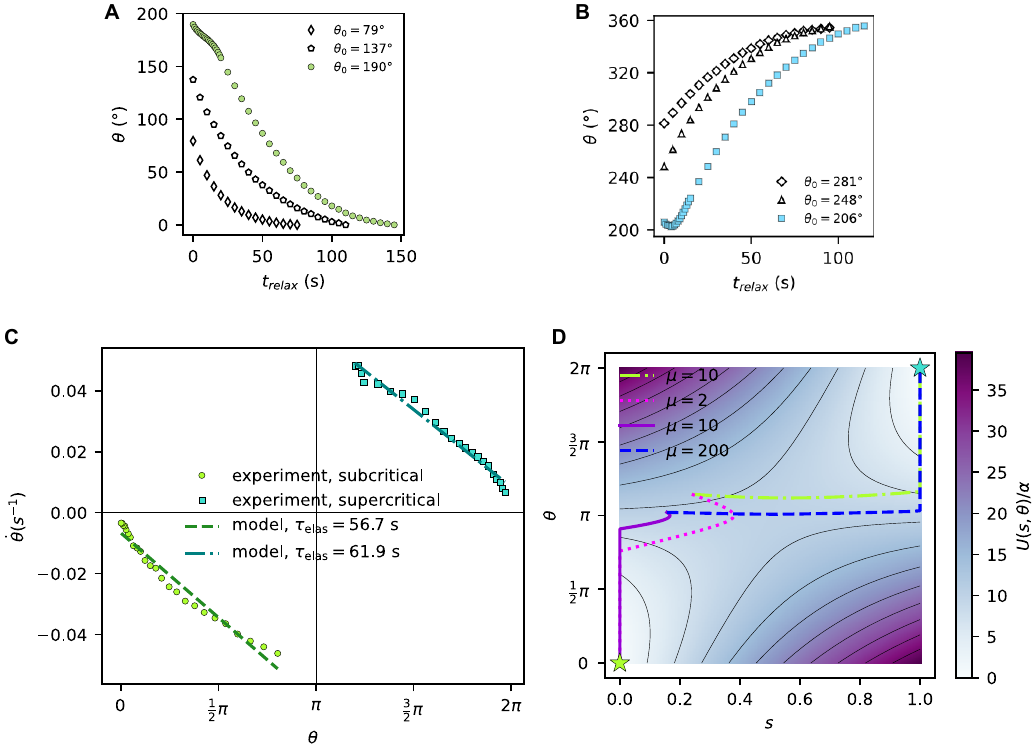}
    \caption{(A) Relaxation of $\theta$ in the subcritical regime for release angle $\theta_0=79^\circ, \ 137^\circ, \ 190^\circ$. (B) Relaxation of $\theta$ in the supercritical regime for release angle $\theta_0=206^\circ, \ 248^\circ, \ 281^\circ$. (C) Linear fit of $\dot{\theta}=-\frac{1}{\tau_\mathrm{elas}} \theta $ for release angle $\theta_0=190^\circ$ (subcritical) and $\theta_0=206^\circ$ (supercritical). (D) Simulated trajectories of relaxation in the elastic energy landscape for two initial points $s_0, \ \theta_0$ and different values of relative drag $\mu=2\pi C_D / C_W$.}
    \label{SI:elastic_relaxation}
\end{figure}

\begin{figure}
    \centering
    \includegraphics[width=1\textwidth]{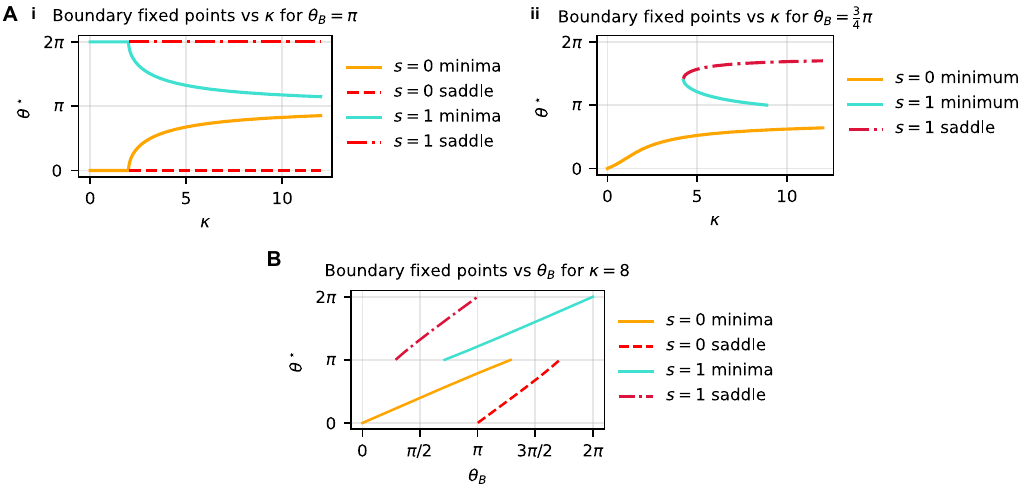}
    \caption{Bifurcation diagrams of the boundary fixed points of the disk-defect system in presence of the magnetic field. (A) i and ii) Bifurcation diagrams vs the relative strength of the magnetic field $\kappa=\mu_B B/\alpha$ for fixed magnetic field orientation $\theta_B = \pi$ and $\frac{3}{4}\pi$, respectively. (B) Bifurcation diagram vs the orientation of the magnetic field $\theta_B$ for fixed magnetic field amplitude $\kappa=8$.}
    \label{SI:bifurcation_diagrams}
\end{figure}

\begin{figure}
    \centering
    \includegraphics[width=1\textwidth]{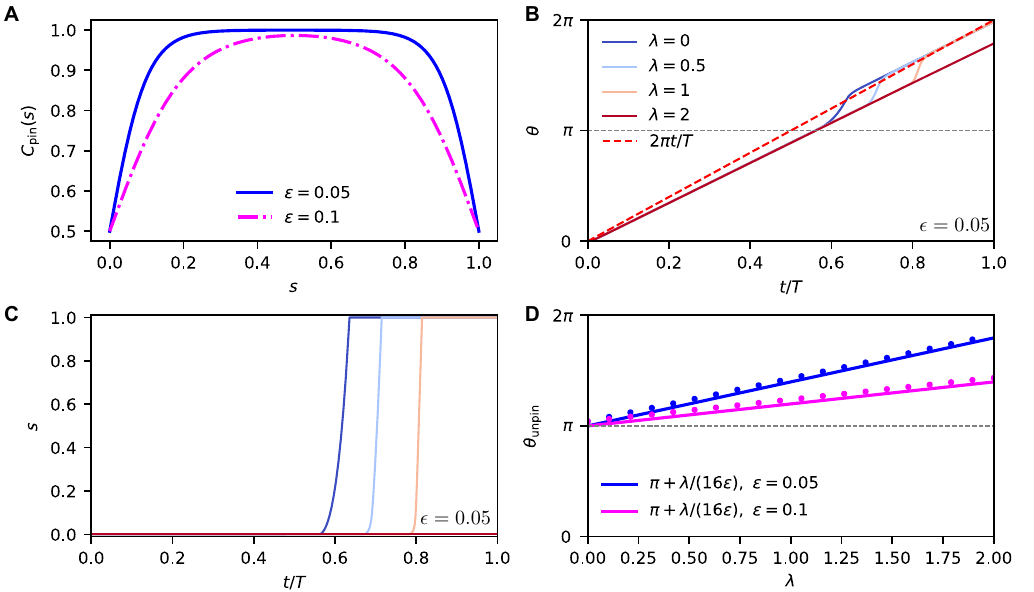}
    \caption{Effect of a pinning energy barrier on defect sweeping during driven disk rotation from the model. (A) Profiles of the pinning energy barrier $C_\mathrm{pin}(s)$ for two values of $\epsilon$, the pinning "slope". (B,C) Simulations of the coupled evolution of $\theta$ and $s$ over one forcing cycle for $\epsilon=0.05$ for increasing pinning energy $\lambda$. (D) The critical angle $\theta_\mathrm{unpin}$ at which sweeping occurs vs $\lambda$ for two values of $\epsilon$ varies as $\pi + \frac{\lambda}{16 \epsilon}$ in simulations.}
    \label{SI:pinning_energy}
\end{figure}

\begin{figure}
\centering
\includegraphics[width=\textwidth]{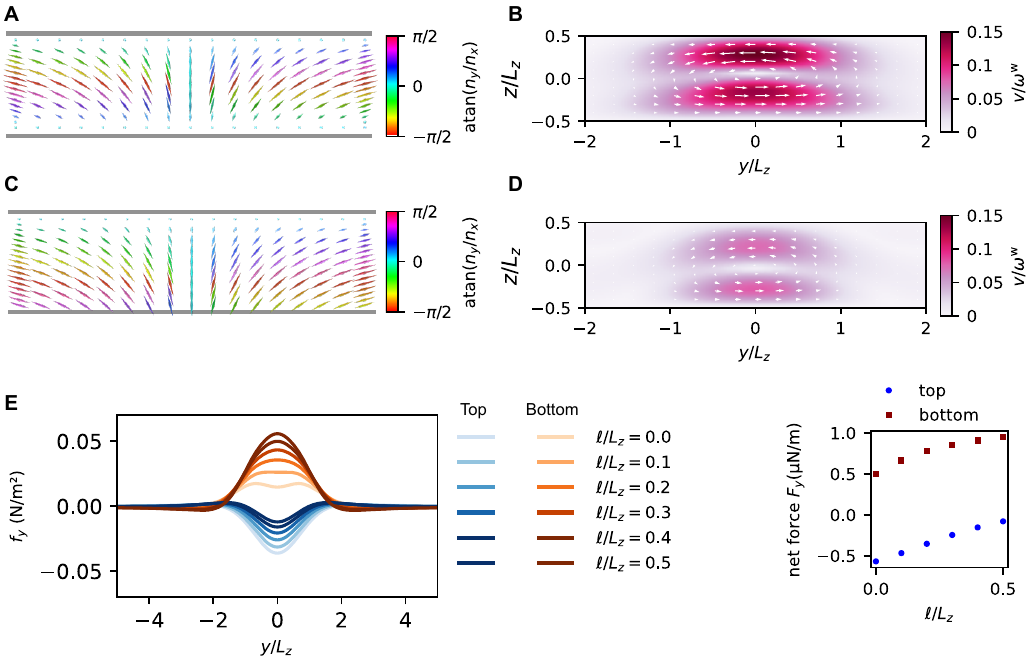}
\caption{Modification of the surface force in case of weak anchoring on the bottom plate. The top plate has infinite anchoring and the bottom plate weak anchoring parameterized by its anchoring extrapolation length $\ell$. We consider a twist wall with $m=-1$ (domain $y<0$ is twisted CCW and domain $y>0$ is twisted CW) and $\epsilon = -1$ (escapes down). (A) director field and (B) velocity field for infinite anchoring with $\ell = 0$µm. (A) director field and (B) velocity field for weak anchoring with $\ell = 5$µm. (E) Horizontal surface force density $f_y$ at the top (blue) and bottom (red) plates. (F) Evolution of the total force on the top and bottom substrates as the anchoring extrapolation length $\ell$ increases from 0 to $L_z/2$.}
\label{SI:hydrodynamics}
\end{figure}

\end{document}